\documentclass{siamltex}
\usepackage{epsfig,latexsym,amssymb}
\usepackage[USenglish]{babel}

\date{\today}
\title{Optimal Covering Tours with Turn Costs\thanks{
An extended abstract version of this paper appears in the {\em Proceedings
of the Twelfth Annual ACM-SIAM Symposium on Discrete Algorithms
(SODA 01)}, 2001, pp.\,138--147~\cite{abdfms-octtc-01}.}
}
\author{
  Esther M. Arkin\footnotemark[2]
\and
  Michael A. Bender\footnotemark[3]
\and
  Erik D. Demaine\footnotemark[4]
\and
  S\'andor~P.~Fekete\footnotemark[5]
\and
  Joseph S. B. Mitchell\footnotemark[2]
\and
  Saurabh Sethia\footnotemark[6]
}

{\makeatletter \gdef\fps@figure{!htbp}}

\newtheorem{exam}[theorem]{Example}
\newtheorem{problem}[theorem]{Problem}

\let\epsilon=\varepsilon
\def\avgdir{{\rho}}   % maximum number of directions
\def\maxdeg{{\delta}}     % maximum degree

\newcommand{\old}[1]{{}}

\newcommand{\comment}[1]
{
  \mbox{}\marginpar{\textsl{\fbox{\vbox{\tiny\raggedright\hspace{0pt}#1}}}}
}

\begin{document}
\maketitle

\renewcommand{\thefootnote}{\fnsymbol{footnote}}

\footnotetext[2]{
    Department of Applied Mathematics and Statistics,
    State University of New York, Stony Brook, NY 11794-3600, USA.
    Email: \{\texttt{estie}, \texttt{jsbm}\}\texttt{@ams.sunysb.edu}.}
\footnotetext[3]{
    Department of Computer Science, State University of New York,
    Stony Brook, NY 11794-4400, USA. 
    Email: \texttt{bender@cs.sunysb.edu}.}
\footnotetext[4]{
    Computer Science and Artificial Intelligence Laboratory,
    Massachusetts Institute of Technology, 
    Cambridge, MA 02139, USA.
    Email: \texttt{edemaine@mit.edu}.}
\footnotetext[5]{
    Department of Mathematical Optimization, 
    Braunschweig University of Technology,
    Pockelsstr.~14, 38106 Braunschweig, Germany.
    Email: \texttt{s.fekete@tu-bs.de}.}
\footnotetext[6]{
    SoftJin Infotech Pvt. Ltd., Bangalore, India.
    Email: \texttt{saurabhsethia@gmail.com}.
    (Research performed while at SUNY Stony Brook.)
    }

\renewcommand{\thefootnote}{\arabic{footnote}}

\begin{abstract}
We give the first algorithmic study of a class of ``covering tour''
problems related to the geometric Traveling Salesman Problem: 
Find a polygonal tour for a {\em cutter} so that it sweeps
out a specified region (``pocket''), 
in order to minimize a cost that depends mainly on the number of
{\em turns}.  These problems arise naturally in manufacturing
applications of computational geometry to automatic tool path
generation and automatic inspection systems, as well as arc routing
(``postman'') problems with turn penalties.  We prove 
the NP-completeness of minimum-turn milling and give efficient
approximation algorithms for several natural versions of the problem,
including a polynomial-time approximation scheme based on a novel
adaptation of the $m$-guillotine method.
\end{abstract}



\pagestyle{myheadings}
\thispagestyle{plain}
\markboth{E. M. ARKIN ET AL.}{COVERING TOURS WITH TURN COST}

\section{Introduction}

An important algorithmic problem in manufacturing is to compute
effective paths and tours for covering (``milling'') a given region
(``pocket'') with a cutting tool. The objective is to
find a path or tour along which to
move a prescribed cutter in order that the sweep of the cutter
covers the region, 
removing all of the material from the
pocket, while not ``gouging'' the material that lies outside of the
pocket.  This covering tour or ``lawn mowing'' problem 
\cite{afm-aalmm-00} 
and its variants arise not only in
NC machining applications but also in 
automatic inspection, spray painting/coating operations,
robotic exploration, arc routing, and even mathematical origami.

The majority of research on these geometric covering tour problems as
well as on the underlying arc routing problems in networks has focused
on cost functions based on the lengths of edges.  However, in many
actual routing problems, this cost is dominated by the cost of
switching paths or direction at a junction. A drastic example is given
by fiber-optical networks, where the time to follow an edge is
negligible compared to the cost of changing to a different frequency
at a router.  In the context of NC machining, turns represent an
important component of the objective function, as the cutter may have
to be slowed in anticipation of a turn.  The number of turns 
(``link distance'') also arises naturally as an objective function in
robotic exploration (minimum-link watchman tours) and in various arc
routing problems,  such as 
snow plowing or street sweeping with turn
penalties. 
R.~Klein~\cite{klein} 
has posed the question of
minimizing the number of turns in polygon exploration problems.

In this paper, we address the problem of minimizing the cost of {\em
turns} in a covering tour.  This important aspect of the problem has
been left unexplored so far in the algorithmic community, 
and the arc routing community has examined only heuristics without
performance guarantee, or exact algorithms with exponential running
time. Thus, our study provides important new insights and a better
understanding of the problems arising from turn cost.
We present several new results: 
\begin{enumerate}

\item[(1)] We prove that the covering tour problem with turn costs is
NP-complete, even if the objective is purely to minimize the number of
turns, the pocket is orthogonal (rectilinear), and the cutter must
move axis-parallel.
The hardness of the problem is not apparent, as our problem seemingly
bears a close resemblance to the polynomially solvable Chinese Postman
Problem; see the discussion below.

\item[(2)] We provide a variety of constant-factor approximation
algorithms that efficiently compute covering tours that are nearly
optimal with respect to turn costs in various versions of the problem.
While getting {\em some} $O(1)$-approximation is not difficult for
most problems in this class, through a careful study of the structure 
of the problem, we have developed tools and techniques that enable
significantly stronger approximation results.

One of our main results is a 3.75-approximation for minimum-turn
axis-parallel tours for a unit square cutter that covers an integral
orthogonal polygon, possibly with holes.  Another main result gives a
4/3-approximation for minimum-turn tours in a ``thin'' pocket, as
arises in the arc routing version of our problem.

Table~\ref{approx summary} summarizes our results.
  The term ``coverage''
  indicates the number of times a point is visited, which is of interest
  in several practical applications. This parameter also provides an upper
  bound on the total length.

\item[(3)] We devise a polynomial-time approximation scheme (PTAS) for
the covering tour problem in which the cost is given as a weighted
combination of length and number of turns, i.e., the Euclidean length
plus a constant $C$ times the number of turns.  For an integral orthogonal polygon with $h$
holes and $N$ pixels, the running time is $2^{O(h)}\cdot N^{O(C)}$.  
The PTAS involves
an extension of the $m$-guillotine method, which has previously been
applied to obtain PTAS's in problems involving only {\em length} 
\cite{m-gsaps-99}.
\end{enumerate}

%%\comment{REFEREE 1: Other examples of its application? Don't try
%to be both theory and applied; admit we are theory.} xxx DONE, see below

We should stress that our paper focuses on the graph-theoretic and algorithmic 
aspects of the turn-cost problem; we make no claims of immediate applicability
of our methods for NC machining.

\begin{table*}
\small{
\centerline{
\begin{tabular}{|l|c|c|c|c|c|}
\hline
& Discrete &          \multicolumn{2}{c|}{Thin} & Orthogonal  & Thin \\
& Milling  &          \multicolumn{2}{c|}{Discrete} & Milling & Orthogonal\\
\hline
Section &\ref{sub:disc.gen} & \multicolumn{2}{c|}{\ref{sub:disc.thin}}  &{\ref{sub:nonint}}&\ref{sub:thin} \\
\hline \hline
Cycle cover APX       & $2\maxdeg + \avgdir$ & 4 & 1.5 & 4.5  & 1\\
Tour APX              & $2\maxdeg +\avgdir+2$ & 6 & 3.5 & 6.25 & 4/3\\
Length APX            & $\maxdeg$            & 2    & - & 8    & 4 \\
Max cover             & $\maxdeg$            & $2\avgdir$ &  - & 8    & 4 \\
Time (explicit) & $O(N)$   & $O(N)$ & $O(N^{3})$ & $O(N^{2.376}+n^3)$ & $O(n^{3})$\\
Time (implicit) & n/a &  n/a     & n/a    & $O(n^{2.5}\log N+n^3)$ & $O(n^{3})$\\
\hline
\end{tabular}
}
\bigskip
\centerline{
\begin{tabular}{|l|c|c|c|}
\hline
&\multicolumn{3}{c|}{Integral}  \\
&\multicolumn{3}{c|}{Orthogonal} \\
\hline
Section & \multicolumn{3}{c|}{\ref{sub:int}} \\
\hline \hline
Cycle cover APX       & 10   & 4 & 2.5 \\
Tour APX              & 12   & 6 & 3.75 \\
Length APX            & 4    & 4 &  4 \\
Max cover             & 4    & 4 &  4 \\
Time (explicit)       & $O(N)$ & $O(N^{2.376})$ & $O(N^{2.376}+n^3)$ \\
Time (implicit)       & $O(n\log n)$& $O(n^{2.5}\log N)$& $O(n^{2.5}\log N+n^3)$ \\
\hline
\end{tabular}
}
%\bigskip
%\centerline{
%\begin{tabular}{|l|c|c|}
%\hline
%& {Orthogonal}   & Thin \\
%& {Milling}      & Orthogonal \\
%\hline
%Discussed in Section  &{\ref{sub:nonint}}&\ref{sub:thin}\\
%\hline \hline
%Cycle cover APX       & 4.5  & 1   \\
%Tour APX              & 6.25 & 4/3 \\
%Length APX            & 8    & 4   \\
%Max cover             & 8    & 4   \\
%Time (explicit) & $O(N^{2.376}+n^3)$ & $O(n^{3})$\\
%Time (implicit) & $O(n^{2.5}\log N+n^3)$ & $O(n^{3})$\\
%\hline
%\end{tabular}
%}
}
%\vspace{1ex}
\caption{\label{approx summary}
 Approximation factors achieved by our (polynomial-time) algorithms.
    Rows marked ``APX'' give approximation factors for
the minimum-turn cycle cover (``Cycle cover''),
the minimum-turn covering tour (``Tour''), and 
the simultaneous approximation of length of a covering tour (``Length'').
The row marked ``Max Cover'' indicates the maximum number of times a point is
    visited. The parameter $\maxdeg$ denotes the maximum
    degree in the underlying graph, while the parameter $\avgdir$ is the maximum number of
    directions in the graph. The two rows for running time refer to
    an ``explicit'' description of input pixels and output and a more compact
    ``implicit'' encoding of pixels and output.
    (See Section~2 for more detailed definitions.)
}
\end{table*}

%\comment{REFEREE 1: Table 1.1 needs work! Forward referencing is confusing;
%organize better; need more detailed discussion than what is in the caption.}
%xxx Disagree. How else should we do this??

\subsubsection*{Related Work}

In the CAD community, there is a vast literature on the subject of
automatic tool-path generation; we refer the reader to
Held~\cite{h-cgpm-91} for a survey and for applications of
computational geometry to the problem.  The algorithmic study of the
problem has focused on the problem of minimizing the length of a
milling tour: Arkin, Fekete, and Mitchell~\cite{afm-lp-93,afm-aalmm-00} 
show that the
problem is NP-hard for the case where the mower is a square. 
Constant-factor approximation
algorithms are given in \cite{afm-lp-93,afm-aalmm-00,irt-tcpaa-94},
with the current best factor being a 2.5-approximation for min-length
milling (11/5-approximation for orthogonal simple polygons).  For the
closely related {\em lawn mowing} problem (also known as the
``traveling cameraman problem''~\cite{irt-tcpaa-94}), in which the
covering tour is not constrained to stay within $P$, the best current
approximation factor is $3+\epsilon$ (utilizing PTAS results for
TSP).  Also closely related is the watchman route problem with limited
visibility (or ``$d$-sweeper problem''); 
Ntafos~\cite{n-wrlv-92} provides a 4/3-approximation, and 
Arkin, Fekete, and Mitchell~\cite{afm-aalmm-00}
improve this factor to 6/5.
The problem
is also closely related to the Hamiltonicity problem in grid graphs;
the results of \cite{ul-hcsgg-97} suggest that in {\em simple}
polygons, minimum-length milling may in fact have a polynomial-time
algorithm.

Covering tour problems are related to {\em watchman route} problems
in polygons, which have received considerable attention in terms of both
exact algorithms (for the simple polygon case) and approximation
algorithms (in general); see \cite{m-gspno-00} for a relatively recent survey.
Most relevant to our problem is the prior work on {\em minimum-link}
watchman tours: see \cite{al-mlvpi-93,al-famlv-95,amp-mlwt-03}
for hardness and approximation results,
and \cite{cm-ilbll-98,kkm-llrwt-94} for combinatorial bounds.
However, in these problems the watchman is assumed to see arbitrarily far,
making them distinct from our tour cover problems.

Other algorithmic results on milling include a study of {\em
multiple tool} milling by Arya, Cheng, and Mount~\cite{acm-aamtm-01},
which gives an approximation algorithm for minimum-length tours that use
different size cutters, and the paper of
Arkin, Held, and Smith~\cite{ahs-oprzp-00}, which examines the problem of
minimizing the number of retractions for ``zig-zag'' machining without
``re-milling'', showing that the problem is NP-complete and giving an
$O(1)$-approximation algorithm.

Geometric tour problems with turn costs have been studied by
Aggarwal~et~al.~\cite{ackms-amtsp-97}, who study
the {\em angular-metric TSP}. The objective is to compute a tour
on a set of points, such that the sum of the direction changes
at vertices is minimized: 
For any vertex $v_i$ with incoming edge $v_{i-1}v_i$ and
outgoing edge $v_iv_{i+1}$, 
the change of direction is given by the absolute value of the
angle between $v_{i-1}v_i$ and $v_iv_{i+1}$.
The problem turns out to be NP-hard,
and an $O(\log n)$ approximation is given.
Fekete~\cite{f-gtsp-92} and Fekete and
Woeginger~\cite{fw-artp-97} have studied a variety of {\em
angle-restricted tour} (ART) problems.
Covering problems of a different nature have been studied
by Demaine et al.~\cite{ddm-ffswp-99}, who considered algorithmic 
issues of origami.

In the operations research literature, there has been an extensive
study of arc routing problems, which arise in snow removal,
street cleaning, road gritting, trash collection, meter reading, mail
delivery, etc.; see the surveys of
\cite{ag-arma-95,egl-arpcp-95,egl-arprp-95}.  Arc routing with turn
costs has had considerable attention, as it enables a more
accurate modeling of the true routing costs in many situations.  Most
recently, Clossey, Laporte, and Soriano~\cite{cls-sarpt-00} 
present six heuristic
methods of attacking arc routing with turn penalties, without
resorting to the usual transformation to a TSP problem; however, their
results are purely based on experiments and provide no provable
performance guarantees.  The {\em directed} postman problem in graphs with turn
penalties has been studied by Benavent and
Soler~\cite{bs-drppt-99}, who prove the problem to be (strongly)
NP-hard and provide heuristics (without performance guarantees) and
computational results.  (See also Fern{\'a}ndez's thesis~\cite{s-pragp-95} and
\cite{ms-escpp-98} for computational experience with worst-case
exponential-time exact methods.)

Our covering tour problem is related to the Chinese Postman Problem (CPP),
which can be solved exactly in polynomial time for ``purely''
undirected or purely directed graphs. However, the
turn-weighted CPP is readily seen to be
NP-complete: Hamiltonian cycle in line graphs is NP-complete (contrary
to what is reported in \cite{gj-cigtn-79}; see page 246 of
West~\cite{w-igt-95}), implying that TSP in line graphs is also
NP-complete.  The CPP on graph $G$ with turn costs
at nodes (and zero costs on edges) is equivalent to TSP on the
corresponding line graph, ${\cal L}(G)$, where the cost of an edge in
${\cal L}(G)$ is given by the corresponding turn cost in $G$.  Thus,
the turn-weighted CPP is also NP-complete.

\section{Preliminaries}
\label{prelim}

This section formally defines the problems at hand, and various special cases
of interest.

\subsubsection*{Problem Definitions}

The general \emph{geometric milling problem} is to find a closed curve
(not necessarily simple) whose Minkowski sum with a given tool
(cutter) is precisely a given region $P$ bounded by $n$ edges. In the context of
numerically controlled (NC) machines, this region is usually called
a {\em pocket}.  Subject to this
constraint, we may wish to optimize a variety of objective functions,
such as the length of the tour, or the number of turns in the tour.
We call these problems \emph{minimum-length} and \emph{minimum-turn}
milling, respectively.  While the latter problem is the main focus of
this paper, we are also interested in bicriteria versions of the
problem in which both length and number of turns must be small; 
we also consider the scenario
in which the objective function is given by a 
linear combination of turn cost and distance traveled
(see Section~\ref{sec:ptas}).

%\comment{REFEREE 2: 
        %How significant is the assumption that the curve is closed to your
        %methods?  Could open curves be allowed, perhaps with a modest
        %increase of running time or approximation factors, or is this
        %assumption critical.  (If you have an easy answer, perhaps a couple
        %sentences here or in the Conclusions would be nice.)}
%xxx DONE, see below
%Finding a shortest tour turns into 
%the question of finding a shortest Hamiltonian path when return
%to the origin is not required. Just like for the TSP, our techniques
%for finding a low-turn tour can still be applied by fixing an edge.
%The resulting approximation factors may vary; details are left to the 
%interested reader.

In addition to choices in the objective function, the problem version
depends on the constraints on the tour.  
%\comment{REFEREE 1: The concept of discrete milling is very vague.
%A figure would help.}
%xxx DONE. No figure, but new text.
%\comment{REFEREE 2: In the discrete milling problem, there seems to be no
        %requirement that the graph be planar.  Is this intended?  (Planarity
        %seems to be an essential element of the geometric problem, but
        %after reading the rest of the paper, it seems that planarity plays
        %no significant role.)}
%xxx DONE. 
The most general case arises when considering a tour that has to visit
a discrete set of vertices, connected by a set of edges, with a specified turn 
cost at each vertex to change from one edge to the next.
More precisely, at each vertex,
the tour has the choice of (0) going ``straight'', if there
is one ``collinear'' edge with the one currently used (costing no turn),
(1) turning onto another, non-collinear edge (costing one turn),
or (2) ``U-turning'' back onto the source edge (costing two turns).
Which pairs of the $d$ edges incident to a degree-$d$ vertex
   are considered ``collinear'' is specified by a matching in the
   complete graph $K_d$: three incident edges cannot be ``collinear''.
We call this graph-theoretic abstraction the 
{\em discrete milling problem}, indicating the close relationship
to other graph-theoretic tour optimization problem.
We are able to give a number of approximation algorithms for 
discrete milling that depend on some graph parameters:
$\maxdeg$ denotes the maximum degree of a vertex and
$\avgdir$ denotes the maximum number of distinct ``directions''
coming together at a vertex. (For example, for graphs 
arising from $d$-dimensional grids, these
values are bounded by $2d$ and $d$, respectively.)

A special case of discrete milling arises when dealing with 
``thin'' structures in two- or three-dimensional
space, where the task is to travel all of a given set of ``channels'', 
which are connected at vertices. This resembles a CPP, 
in that it requires us to travel a given set of edges; 
however, in addition to the edge cost, there is a cost at the vertices
when moving from one edge to the next. For this scenario, we are able
to describe approximation factors that are independent of other graph 
parameters.
%\comment{REFEREE 1: Thin milling -- why is it important?  Connection
%to CPP not clear!}
%xxx DONE.

\begin{figure}[htbp]
\begin{center}
\epsfig{figure=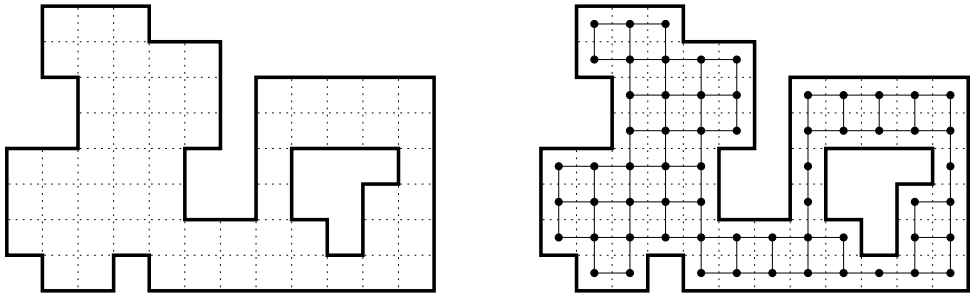,width=0.99\linewidth}
\end{center}
\caption{\label{integral orthogonal}
  An instance of the integral orthogonal milling problem (left)
  and the grid graph model (right).}
\end{figure}

More geometric problems arise when considering the milling of a polygonal
region $P$. In the \emph{orthogonal
milling problem}, the region $P$ is an orthogonal polygonal domain
(with holes) and the tool is an (axis-parallel) unit-square cutter
constrained to axis-parallel motion, with edges of the tour alternating between
horizontal and vertical.  All turns are orthogonal;
$90^\circ$ turns incur a cost of 1, while a ``U-turn'' has a cost of~2.
In the {\em integral orthogonal} case, all coordinates of boundary edges
are integers, so the region can be considered to be the (connected) union
of $N$ {\em pixels}, i.e., axis-parallel unit squares with integer vertices.
Note that in general, $N$ may not be bounded by a polynomial in $n$.
Instead of dealing directly with a geometric milling problem, we often
find it helpful to consider a more combinatorial problem, and then
adapt the solution back to the geometric problem.  
In particular, for integral orthogonal milling, we may assume that 
an optimal tour can be assumed to have its vertex
coordinates of the form $k+\frac{1}{2}$ for integral $k$.  Then,
milling in an integral orthogonal polygon (with holes) is equivalent
to finding a tour of all the vertices (``pixels'') of a grid graph; see Figure
\ref{integral orthogonal}.  

An interesting special case of integral orthogonal milling
is the \emph{thin orthogonal milling problem}, in which 
the region does not contain a 2$\times$2 square
of pixels. This is 
also closely related to discrete milling, as
we can think of edges embedded
into the planar grid, such that vertices and channels are well separated.
This problem of finding a tour with minimum turn cost for this
class of graphs is still NP-complete, even for a subclass
for which the corresponding problem of minimizing total distance
is trivial; this highlights
the particular difficulty of dealing with turn cost.
On the other hand, thin orthogonal milling allows for particularly
fast and efficient approximation algorithms.

\subsubsection*{Other Issues}

\begin{figure}[htb]
\leavevmode
\begin{center}
\leavevmode
\epsfig{file=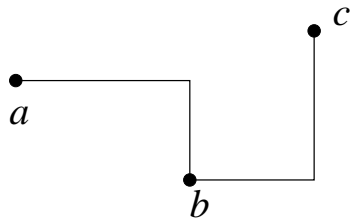,width=.35\textwidth} 
\end{center}
\caption{The triangle inequality may not hold when using turn cost 
as distance measure: $d(a,c)=3>2=d(a,b)+d(b,c)$}
\label{fi:no.triangle}
\end{figure}

It should be stressed that using turn cost instead of (or in addition
to) edge length changes several characteristics of distances. One
fundamental problem is illustrated by the example in
Figure~\ref{fi:no.triangle}: the triangle inequality does not have to
hold when using turn cost. This implies that many classical
algorithmic approaches for graphs with nonnegative edge weights (such
as using optimal 2-factors or the Christofides method for the TSP)
cannot be applied without developing additional tools.

In fact, in the presence of turn costs we distinguish between the
terms {\em 2-factor}, i.e., a set of edges, such that
every vertex is incident to two of them,
and {\em cycle cover}, i.e., a set of cycles, such that every
vertex is covered. While the terms are
interchangeable when referring to the set of edges that they
constitute, we make a distinction between their respective {\em
costs}: a ``2-factor'' has a cost consisting of the sum of edge costs,
but does not necessarily account for the turn cost between it two
incident edges, while the cost of a ``cycle cover'' includes 
also the turn costs at vertices.

It is often useful in designing approximation algorithms for optimal
{\em tours} to begin with the problem of computing an optimal
\emph{cycle cover}, minimizing the total number of turns in a set of
cycles that covers $P$.  Specifically, we can decompose the problem of
finding an optimal (minimum-turn) tour into two tasks: finding an
optimal cycle cover, and merging the components.  Of course, these two
processes may influence each other: there may be several optimal cycle
covers, some of which are easier to merge than others.  (In
particular, we say that a cycle cover is {\em connected}, if the graph
induced by the set of cycles and their intersections is connected.)
As we will show, even the problem of optimally merging a connected
cycle cover is NP-complete. This is in contrast to minimum-length
milling, where an optimal connected cycle cover can trivially be
converted into an optimal tour that has the same cost.

Algorithms whose running time is polynomial in the explicit encoding size
(pixel count) are \emph{pseudo-polynomial}.  Algorithms whose running
time is polynomial in the implicit encoding size are \emph{polynomial}.
This distinction becomes an
important issue when considering different ways to encode input and output;
e.g., a large set of pixels forming an $a\times b$ rectangle 
can be described in space $O(\log a + \log b)$ by simply describing the 
bounding edges, instead of listing all $ab$ individual pixels.
In integral orthogonal milling, one might think that it is most natural
to encode the grid graph with vertices, 
because the tour will be embedded on this
graph and will, in general, have complexity proportional to the number $N$
of pixels.  But the input to any geometric milling problem has a
natural encoding by specifying only the $n$ vertices of the polygon $P$.  In
particular, long edges are encoded in binary (or with one real number,
depending on the model) instead of unary.  It is possible to get a
running time depending only on this size, but of course we need to
allow for the output to be encoded \emph{implicitly}.  That is, we
cannot explicitly encode each vertex of the tour, because there are
too many (the number can be arbitrarily large even for a succinctly encodable
rectangle).  Instead, we encode an abstract description of the tour
that is easily decoded. 

Finally, we mention that many of our results carry over from the {\em
tour} (or cycle) version to the {\em path} version, in which the
cutter need not return to its original position. In this paper, 
we omit the straightforward 
changes necessary to compute optimal paths. A similar adjustment 
can be made for the related case
of {\em lawn mowing}, in which the sweep of the cutter is allowed to
go outside $P$ during its motion. Clearly, our techniques are also useful for
scenarios of this type.

\section{NP-Completeness}
\label{NP Completeness}

Arkin, Fekete, and Mitchell~\cite{afm-aalmm-00} have proved that the problem of
optimizing the length of a milling tour is NP-hard. Their proof
is based on the well-known hardness of deciding whether a 
grid graph has a Hamiltonian cycle \cite{ips-hpgg-82,jp-cctsp-85}.
This result implies that it is NP-hard to find a tour of minimum total
length that visits all vertices. If, on the other hand, we are given 
a connected cycle cover of a graph that has minimum
total length, then it is trivial to convert it into a tour of 
the same length by merging the cycles into one tour.

In this section we show that if the quality of a tour is measured by
counting {\em turns}, then even this last step of turning an optimal
connected cycle cover into an optimal tour is NP-complete. Thus
we prove that it is
NP-hard to find a milling tour that optimizes the
number of turns for a polygon with holes.

\begin{theorem} \label{thin hard}
Minimum-turn milling is NP-complete, even when we are
restricted to the orthogonal thin case, and are
already provided with an optimal connected cycle cover.
\end{theorem}

Because thin orthogonal milling is a special case of thin milling as well as
orthogonal milling, and it is easy to convert an instance of thin orthogonal
milling into an instance of integral orthogonal milling, we have

\begin{corollary}
Discrete milling, 
orthogonal milling,
and integral orthogonal milling are NP-complete.
\end{corollary}

\begin{proof}[Theorem \ref{thin hard}]
Our reduction proceeds in two steps. First we show that
the problem {\sc Hamiltonicity of Unit Segment Intersection Graphs}
({\sc Husig}) of deciding the Hamiltonicity of 
intersection graphs of axis-parallel unit segments is hard.
To see this, we use the NP-hardness of deciding Hamiltonicity of grid graphs (\cite{ips-hpgg-82,jp-cctsp-85}) and
argue that any grid graph can be represented in this form; see Figure~\ref{fi:intersect}:

\begin{figure}[htbp]
\begin{center}
\epsfig{figure=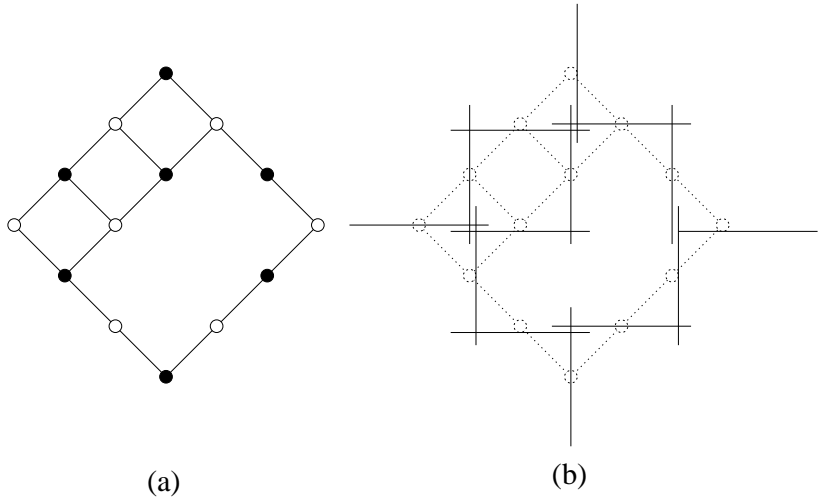,width=.83\linewidth}
\end{center}
\caption{(a) A grid graph $G$. (b) A representation of $G$
as an intersection graph of axis-parallel unit segments.}
\label{fi:intersect}
\end{figure}

Consider a set of integer grid points that induce a grid graph $G$.
Note that $G$ is bipartite, because one can 2-color the
nodes by coloring a grid point $(x,y)$ 
black (resp., white) if $x+y$ is odd (resp., even).
After rotating the point set
by $\pi/4$, the coordinate of each point is an integer multiple
of $1/\sqrt{2}$. Scaling down the resulting arrangement
by a factor of $3/\sqrt{2}$ results in an arrangement
in which the coordinate of each point is an integer
multiple of $1/3$, and the shortest distance between two points
of the same color class is $2/3$. 
For the resulting set of points $p_i=(x_i,y_i)$,
let $p'_i=p_i+(\epsilon_i,\epsilon_i)$
be given as the set obtained by ``perturbations''
$\epsilon_i$ that are small and all distinct.
Then, represent each ``white'' vertex
by a horizontal unit segment centered at 
$p'_i$, and each ``black'' vertex
by a vertical unit segment centered at $p'_i$.
Now it is easy to see that the resulting
unit segment intersection graph is precisely the original grid graph~$G$.

In a second step, we show that the problem {\sc Husig} reduces
to the problem of milling with turn costs.
The outline of our argument is illustrated in Figure~\ref{fi:thin.hard}.

%\comment{REFEREE 1: Figure 3.2 hard to follow!}
% xxx DONE. Labeled figure and rewrote proof to include more details.

\begin{figure}[htbp]
\begin{center}
\epsfig{figure=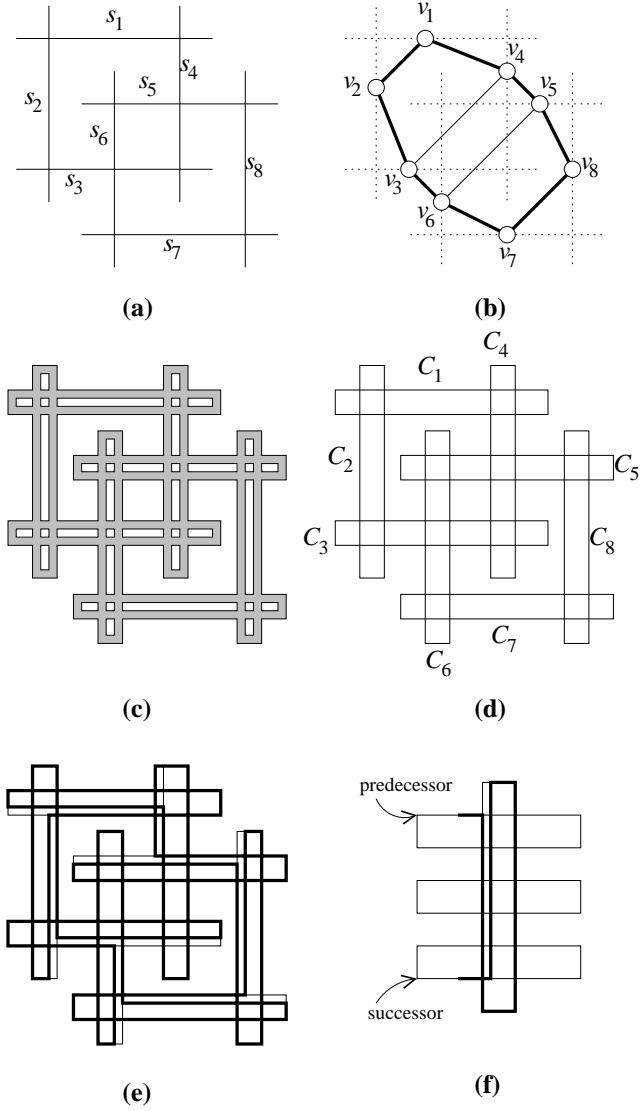,width=.65\linewidth}
\end{center}
\caption{Thin orthogonal milling with turn cost is NP-hard:
(a) a set of $s=8$ axis-parallel unit segments, denoted by $s_1,\ldots,s_8$;
(b) the corresponding intersection graph $G$, with the Hamiltonian
cycle $v_1,v_2,v_3,v_6,v_7,v_8,v_5,v_4$ shown in bold;
(c) representing $G$ by a connected region consisting of $4s$ corridors; 
(d) a drawing of the graph induced by the instance of thin orthogonal milling, with the
$s=8$ rectangular cycles $C_1,\ldots,C_8$;
(e) a milling tour with $5s$ turns corresponding to the Hamiltonian cycle in 
$G$;
(f) milling the four corridors of a cycle using five turns.}
\label{fi:thin.hard}
\end{figure}

Consider a unit segment intersection graph $G$, given by a set
of axis-parallel unit segments, as shown in
Figure~\ref{fi:thin.hard}(a).  Figure~\ref{fi:thin.hard}(b) shows
the corresponding graph, with a Hamiltonian cycle indicated
in bold. Without loss of generality, we may
assume that $G$ is connected. Let $s$ be the number of nodes of $G$.

As shown in Figure~\ref{fi:thin.hard}(c), 
we replace each line segment by a cycle of 
four thin axis-parallel
corridors. This results in a connected polygonal region $P$ having $4s$
convex corners. Clearly, any cycle
cover or tour cover of $P$ must have at least $4s$ turns; by using a cycle
for each set of four corridors representing a strip,
we get a cycle cover $\cal C$ with $4s$ turns. Therefore, $\cal C$
is an optimal cycle cover, and it is connected, because $G$ is connected.

Now assume that $G$ has a Hamiltonian cycle. It is easy to see
(Figure~\ref{fi:thin.hard}(f)) that this cycle can be used to
construct a milling tour of $P$ with a total of $5s$ turns: Each time
the Hamiltonian cycle moves from one vertex $v_i$ of the grid graph to the next
vertex $v_j$, the milling tour moves from the cycle $C_i$
representing $v_i$ to the cycle $C_j$ representing $v_j$, 
at an additional cost
of 1 turn for each of the $s$ edges in the Hamiltonian cycle.

Assume conversely that there is a milling tour $T$ with at most $5s$
turns. We refer to turns at the corners of 4-cycles as {\em convex
turns}. The other turns are called {\em crossing turns}.

As noted above, the convex corners of $P$ require at least $4s$ convex
turns. Consider the sequence of turns $t_1,\ldots,t_{5s}$ in $T$. By
construction, the longest contiguous subsequence of convex turns
contains at most four different convex turns.  (More precisely, we can
only have such a subsequence with four different convex corners, if
these four corners belong to the same 4-cycle representing a unit
segment.) Furthermore, we need at least one additional crossing turn
at an interior crossing of two corridors to get from one convex corner
to another convex corner not on the same 4-cycle.  (More precisely,
one crossing turn is sufficient only if the two connected convex
corners belong to 4-cycles representing intersecting unit segments.)
Therefore, we need at least $c$ crossing turns if we have at least $c$
contiguous subsequences as described above.  This means that $c\geq
s$; hence, $c=s$ by the assumption on the number of turns on $T$. Because the
$c$ crossing turns correspond to a closed roundtrip in $G$ that visits
all $s$ vertices, this implies that we have a Hamiltonian cycle,
concluding the proof.
\hfill\end{proof}

\section{Approximation Tools}
\label{sec:tools}

There are three main tools that we use to develop approximation
algorithms: computing optimal cycle covers for milling the
``boundary'' of $P$ (Section \ref{ssec:boundary}), converting
cycle covers into tours (Section \ref{Merging Cycles}), and using
optimal (or nearly-optimal) ``strip covers'' (Section \ref{Strip
Cover}). In this section, our description mostly focuses on orthogonal milling;
however, we will see in the following Section~5.1 how some of our tools
can also be applied to the general case of discrete milling.

\subsection{Boundary Cycle Covers}
\label{ssec:boundary}
%\comment{SANDOR: I rewrote Section 4.1. Please proofread!}
We consider first the problem of finding a minimum-turn cycle cover for 
covering a certain subset, $\overline P$, of $P$ that is along its boundary.  
This will turn out to be a useful tool for approximation algorithms.
Specifically, in orthogonal milling we define the set $\overline{P}$
of \emph{boundary pixels} to consist of
pixels that have at least one of their four edges on a boundary
edge of the polygon; i.e., in the grid graph that describes adjacency
of pixels, these are pixels of degree  at most 3. 
Let $N_{\overline{P}}$ be the number of boundary pixels.
A {\em boundary cycle cover} is a collection of 
cycles that visit all boundary pixels.

\begin{figure}[htb]
\leavevmode
\begin{center}
\leavevmode
\epsfig{file=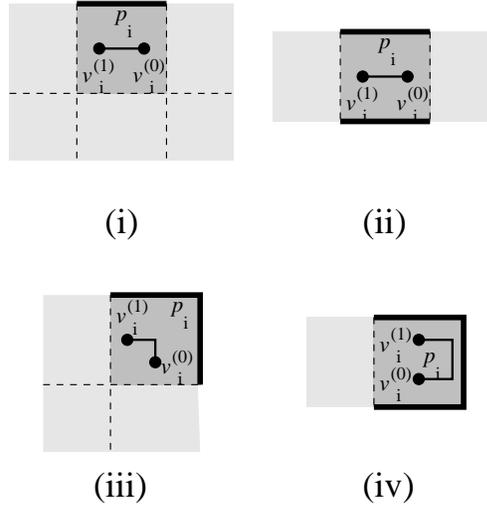,width=.5\linewidth} 
\end{center}
\caption{Representing a boundary pixel $p_i$ by a pair of vertices $v_i^{(0)}$
and $v_i^{(1)}$.}
\label{fig:boundary.graph}
\end{figure}

We define an auxiliary structure, $G_{\overline{P}}=(V_{\overline{P}},E_{\overline{P}})$, which is
a complete weighted graph on $2N_{\overline{P}}$ vertices; for ease of
description, we will refer to $G_{\overline{P}}$ as a set of points
and paths between them. This will allow
us to map boundary cycle covers in $P$ to matchings of corresponding
turn cost in $G_{\overline{P}}$. For this purpose, map each pixel $p_i\in \overline{P}$
to two vertices in $V_{\overline{P}}$, $v_i^{(0)}$ and $v_i^{(1)}$. For each
boundary pixel $p_i$, this pair represents an orientation that is 
attained by a cutter when visiting $p_i$. Depending on the boundary
structure of $p_i$, there are four different cases; 
refer to Figure~\ref{fig:boundary.graph}.

{\bf (i)} One edge of $p_i$ is a boundary edge of the polygon.

{\bf (ii)} Two opposite edges of $p_i$ are boundary edges of the polygon.

{\bf (iii)} Two adjacent edges of $p_i$ are boundary edges of the polygon.

{\bf (iv)} Three edges of $P_i$ are boundary edges of the polygon.

For easier description, we refer to the vertices 
$v_i^{(0)}$ and $v_i^{(1)}$ as points embedded within $p_i$, as shown
in Figure~\ref{fig:boundary.graph}. Furthermore, we add
a {\em mandatory path} $m_i$ between $v_i^{(0)}$ and $v_i^{(1)}$,
represented by a polygonal path with $c(m_i)=0$ (cases (i) and (ii)), 
$c(m_i)=1$ (case (iii)) or $c(m_i)=2$ turns (case (iv)), as shown in the figure.
This path maps the contour of $P$ at $p_i$, and it
represents orientations that a cutter has to attain
when visiting pixel $p_i$.
Note that traveling from $v_i^{(h)}$ to $v_j^{(1-h)}$ along $m_i$
induces a heading $\delta_i^{(h,-)}$ when leaving $v_i^{(h)}$ and 
a heading $\delta_i^{(1-h,+)}$ when arriving at $v_i^{(1-h)}$.
Note that $\delta_i^{(h,-)}$ is opposite to $\delta_i^{(h,+)}$.

\begin{figure}[htb]
\leavevmode
\begin{center}
\leavevmode
\epsfig{file=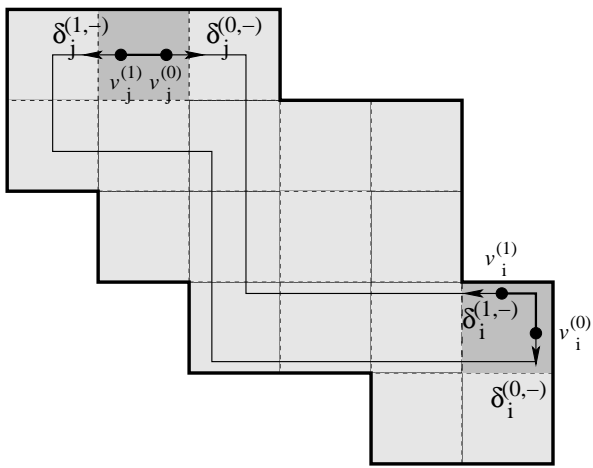,width=.6\linewidth} 
\end{center}
\caption{The cost of traveling between two pixels:
$d(v_i^{(1)},v_j^{(0)})=d(v_j^{(0)},v_i^{(1)})=2$, while
$d(v_i^{(0)},v_j^{(1)})=d(v_j^{(1)},v_i^{(0)})=5$.
}
\label{fig:heading}
\end{figure}

Now we add a set of {\em optional paths}, representing the weighted
edges $E_{\overline{P}}$ of the complete graph $G_{\overline{P}}$. For an example,
refer to Figure~\ref{fig:heading}.
For any pair of vertices
$v_i^{(h)}$ and $v_j^{(k)}$, 
let $d(v_i^{(h)},v_j^{(k)})$ be the minimum
number of turns necessary  when traveling from $v_i^{(h)}$ with
heading $\delta_i^{(h,+)}$ to $v_j^{(k)}$ with heading
$\delta_j^{(k,-)}$.
Note that $d(v_i^{(h)},v_j^{(k)})=d(v_j^{(k)},v_i^{(h)})$,
as any shortest path can be traveled in the opposite direction.
Using a Dijkstra-like approach, we 
can compute these distances
from one boundary pixel to all 
other boundary pixels in time $O(N_{\overline{P}}\log N_{\overline{P}})$; see the overview 
in \cite{m-gspno-00} or the paper \cite{m-lsppo-92}.
The overall time of $O(N_{\overline{P}}^2\log N_{\overline{P}})$
for computing all these link distances is dominated by the following step:
In time $O(N_{\overline{P}}^{3})$ \cite{g-iammng-73,s-co-03}, 
find a minimum-weight perfect matching in the
complete weighted graph $G_{\overline{P}}$.

Now it is not hard to see the following.

\begin{lemma}
\label{le:map}
Any boundary cycle cover in $P$ with $t$ turns can be mapped to
a perfect matching in $G_{\overline{P}}$ of cost $t-\sum_{p_i\in\overline{P}}c(m_i)$,
and vice versa.
\end{lemma}

\begin{proof}
Whenever a boundary cycle cover visits a boundary pixel $p_i$, it has to
perform the turns corresponding to the mandatory path $m_i$. Moreover,
moving from one pixel $p_i$ to the next pixel $p_j$
can be mapped to the optional path
corresponding to the edges $(v_i^{(h)},v_j^{(k)})$; 
clearly, the overall cost is as stated.

Conversely, it is straightforward to see that the combination of a 
perfect matching in $G_{\overline{P}}$ and the mandatory paths yields a boundary
cycle cover of corresponding cost.
\hfill\end{proof}

%,offsets of each boundary edge, by $0.5$ towards the interior of $P$,
%shrunk by $0.5$ at each end.
%More generally, if the tool is such that it can sweep ``against''
%any edge of the polygon and cover a small region adjacent to that
%edge, we define these sweeping line segments to be the boundary edges.
%The region $\overline P$ is defined, then, to be the Minkowski sum of
%the boundary edges and the tool, and we say that a cycle cover or tour
%\emph{mills the boundary} if it covers $\overline P$.  
%For the orthogonal case, we 
%exploit the following property:

%\comment{REFEREE 1: clarify Lemma 4.1!
%Distinguish between ``polygon edge'' and ``tour edge'';
%``boundary edge covered by a polygon edge''??}
%xxx DONE.

 Using the algorithms described above, we also obtain the following:

\begin{theorem} 
\label{th:boundary}
Given the set of $N_{\overline{P}}$ boundary pixels, a minimum-turn 
boundary cycle cover can be computed in time
$O(N_{\overline{P}}^3)$, the time it takes to compute a perfect matching
in $G_{\overline{P}}$.
\end{theorem}

%\comment{REFEREE 1: Figure 4.1 hard to follow!}
%xxx DONE.

If the set of pixels is not given in unary, but implicitly as the
pixels contained in a region with $n$ edges, the above complexity
is insufficient. However, we can use local modifications
to argue the following tool for speeding up the search for an
optimal perfect matching.

\begin{lemma} \label{le:local}
Let $p_i$ and $p_j$ be neighboring boundary pixels that are adjacent to the 
same boundary edge, so $d(v_i^{(h)},v_j^{(k)})=0$ for an appropriate
choice of $h$ and $k$. Then there is an optimal matching
containing $(v_i^{(h)},v_j^{(k)})$.
\end{lemma}

\begin{figure}[htb]
\leavevmode
\begin{center}
\leavevmode
\epsfig{file=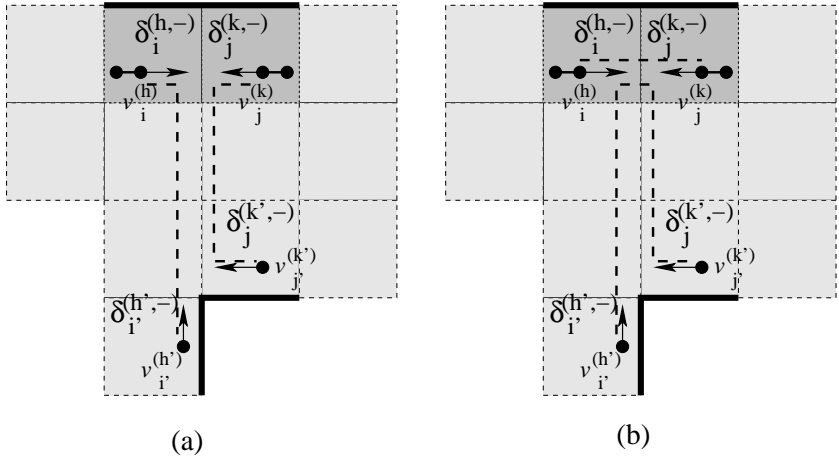,width=.85\linewidth} 
\end{center}
\caption{By performing local modifications, an optimal cycle cover
can be assumed to cover each collinear piece of the boundary in 
one connected strip.}
\label{fig:local}
\end{figure}

\begin{proof}
This follows by a simple exchange argument. See Figure~\ref{fig:local}.
Suppose two adjacent pixels $p_i$ and $p_j$ 
along the same boundary edge are not matched to
each other, 
let $v_i^{(h)}$ be the vertex such that
$\delta_i(h,-)$ is heading for $p_j$, and
let $v_j^{(k)}$ be the vertex such that
$\delta_j(k,-)$ is heading for $p_i$. Furthermore suppose
that $v_{i'}^{(h')}$ is matched to $v_i^{(h)}$ and 
$v_{j'}^{(k')}$ is matched to $v_j^{(k)}$. Then we can match
$v_{i'}^{(h')}$ to $v_{j'}^{(k')}$ and $v_i^{(h)}$ to $v_j^{(k)}$
without changing the cost of the matching.
\hfill\end{proof}

This allows us to obtain a strongly polynomial version
of the matching algorithm of Theorem~\ref{th:boundary}.

\begin{theorem} \label{th:strong}
A minimum-turn boundary cycle cover can be computed in time
$O(n^3)$. 
\end{theorem}

\begin{proof}
By applying Lemma~\ref{le:local} repeatedly, we get $O(n)$ connected 
boundary strips, consisting of sets of collinear boundary
pixels. These can be determined efficiently by computing
offsets of the boundary edges. This leaves only $O(n)$ endpoints
of such strips to be matched, resulting in the claimed complexity.
\hfill\end{proof}

%Suppose that we have a boundary edge
%that is not entirely covered by some tour edge. This means that the remaining
%portion of the edge must be covered by some other edge. This allows
%us to do a local modification that does not increase the number
%of (weighted) turns, but extends the part of the boundary edge that
%is covered by the same edge. (Refer to Figure \ref{fi:local}.)

Note that the validity of this
argument is not restricted to the {\em integral} orthogonal case,
but remains valid even for orthogonal regions with arbitrary boundary
edges.
%\hfill\end{proof}

%This property allows us to apply methods similar to those used in
%solving the CPP: we know portions of edges that
%must be in the cycle cover, and furthermore these edges mill the
%boundary.  What remains is to connect these edges into cycles, while
%minimizing the number of additional turns.
%
%The crucial knowledge that we are using is the
%orientation of the boundary edges, so, for example, we correctly
%compute that the turn distance is zero when two boundary edges are
%collinear. 
%graph on boundary-edge endpoints, with each edge weighted
%according to the corresponding turn distance.  This connects the
%boundary edges optimally into a set of cycles.  
%%(Of course, there may be more
%than one cycle.)  This proves

%\begin{theorem} \label{boundary cycle cover}
%A minimum-turn cycle cover of the boundary of a region can be computed
%in time $O(n^{3})$, which is 
%the time for computing a minimum-weight perfect matching.
%\end{theorem}

\begin{figure}[htbp]
\begin{center}
\epsfig{figure=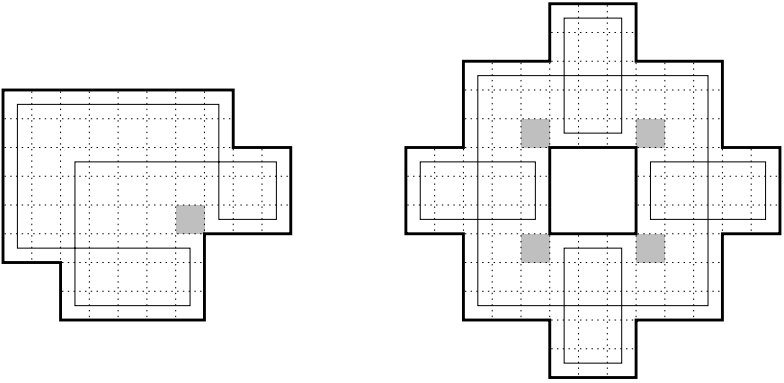,width=0.8\linewidth}
\end{center}
\caption{\label{nasty boundary}
  Optimally covering all pixels that have an edge against the boundary
  can leave reflex pixels uncovered.}
\end{figure}

\paragraph{Remark}
The definition of the ``boundary'' pixels $\overline P$ used
here does not include all pixels that touch the boundary of $P$
in a diagonal fashion; in
particular, it omits the ``reflex pixels'' that share a corner, but no
edge, with the boundary of $P$.  It seems difficult to require that
the cycle cover mill reflex pixels, because Lemma \ref{le:local}
does not extend to this case, and an optimal cycle cover of the
boundary (as defined above) may have fewer turns than an optimal cycle
cover that mills the boundary $\overline P$ plus the reflex pixels;
see Figure \ref{nasty boundary}.

\subsection{Merging Cycles}
\label{Merging Cycles}

It is often easier to find a minimum-turn cycle cover (or
constant-factor approximation thereof) than to find a minimum-turn
tour. 
We show that an exact or approximate minimum-turn cycle
cover implies an approximation for a minimum-turn tour.

We concentrate on the integral orthogonal case.
First we define a few terms precisely.
Two pixels are \emph{adjacent} if the distance between their centers
is $1$.  Two cycles $T_1$, $T_2$ are \emph{intersecting} iff 
$T_1 \cap T_2 \neq \emptyset$.
Two cycles are called \emph{touching} if and only if they are not intersecting and there
exist pixels $p_1 \in T_1$, $p_2 \in T_2$ such that $p_1$ and $p_2$
are adjacent.

\begin{lemma} \label{merge-intersecting}
Let $P_1$ and $P_2$ be two cycles, with $t_1$ and $t_2$ turns, respectively,  
and let $p$ be a pixel that is contained in both cycles. 
Then there is a cycle milling the union of pixels milled by $P_1$ and
$P_2$ and having at most $t_1 + t_2 + 2$ turns.  This cycle can be found in
time linear in the number of its turns.
\end{lemma}

\begin{proof}
Let the neighbors of $p$ in $P_1$ be $a_1$, $a_2$ 
and those of $p$ in $P_2$ be $b_1$, $b_2$.
Connect $a_1$ via $p$ to $b_1$ and $a_2$ via $p$ to $b_2$ 
to get the required tour.  The two
connections may add at most a turn each.  Hence the resulting tour can be of
size at most $t_1 + t_2 + 2$.
\hfill\end{proof}

\begin{lemma} \label{merge-touching}
Given two touching cycles $T_1$, $T_2$ with $t_1$, $t_2$ turns respectively,
then there is a tour $T$ with at most $t_1 + t_2 + 2$ turns
that mills the union of pixels milled by $T_1$, $T_2$.
\end{lemma}

\begin{figure}[htbp]
\begin{center}
\epsfig{figure=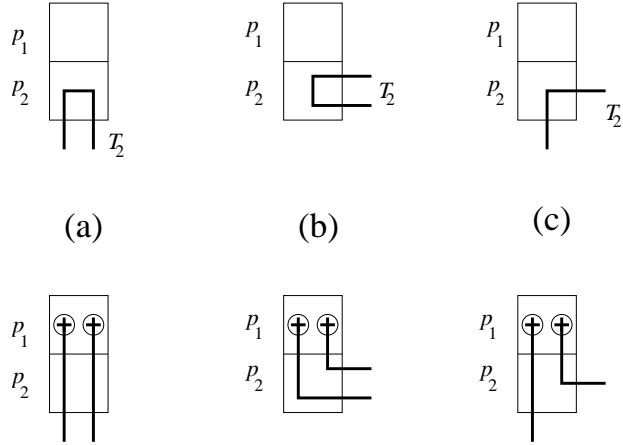,width=3.2in}
\end{center}
\caption{Merging two touching tours: There are three possible ways 
of tour $T_2$ visiting pixel $p_2$ (above). In each case, we can modify $T_2$
into a path that visits pixel $p_1$ (below); at a cost of possibly one extra turn
at each end of the path, we can merge it with tour $T_1$.}
\label{three-ways}
\end{figure}

\begin{proof}
Because $T_1$, $T_2$ are touching, $T_1 \cap T_2 = \emptyset$ and there
exist adjacent pixels $p_1 \in T_1$ and $p_2 \in T_2$.  Without loss
of generality assume that $p_2$ is a leftmost such pixel,
and below $p_1$.
Due to these constraints, $T_2$ can enter/exit $p_2$ from only two sides.
Hence there are only three ways in which $T_2$ can visit $p_2$.  These are
shown in Figure~\ref{three-ways}.  For all three ways we show in
Figure~\ref{three-ways} how to cut and extend tour $T_2$
without adding any extra turns, to get a path $P_2$ starting
and ending at pixel $p_1$.  Cut $T_1$ at $p_1$ to get a path $P_1$. 
By possibly adding two turns, we can merge the two paths into one tour. 
\hfill\end{proof}

With the help of these lemmas, we deduce the following:

\begin{theorem} \label{merging cycles}
A cycle cover with $t$ turns can be converted into a tour with at most 
$t+2(c-1)$ turns, where $c$ is the number of cycles.
\end{theorem}

\begin{proof}
We prove this theorem by induction on the number of tours, $c$, in
the cycle cover.  The theorem is trivially true for $c = 1$.
For any other $c$, choose any $c - 1$ cycles with a total of $t^\prime$
turns and find a tour $T^\prime$ that covers those $c - 1$ cycles; by
induction, it has $t^\prime + 2(c - 1)$ turns.
Let the remaining cycle, $R$, have $r$ turns.  Thus $t = t^\prime + r$.
Because the polygon is connected, the set of pixels milled by $R$ and $T^\prime$
must be connected.  Hence either $T^\prime$ and $R$ are intersecting or
touching.  By Lemmas~\ref{merge-intersecting} and ~\ref{merge-touching} we
can merge $R$ and $T^\prime$ into a single tour $T$ with at most
$t^\prime + 2(c - 1) + r + 2$ turns, i.e. $t + 2c$ turns.
\hfill\end{proof}

\begin{corollary} \label{cycle-cover-merging}
A cycle cover of a connected rectilinear polygon with $t$ turns can
be converted into a single milling tour with at most $\frac{3}{2} t$ turns.
\end{corollary}

\begin{proof}
This follows immediately from Theorem~\ref{merging cycles} and the fact that
each cycle has at least four turns.
\hfill\end{proof}

Unfortunately, general merging is difficult (as illustrated by the 
NP-hardness proof of Theorem~\ref{thin hard}), so we cannot hope to 
improve these general merging results by more than a constant factor.

\subsection{Strip and Star Covers}
\label{Strip Cover}

A key tool for approximation algorithms is a covering of the region by a
collection of ``strips.''  A \emph{strip} is a maximal straight
segment whose
Minkowski sum with the tool is contained in the region.  A \emph{strip cover}
is a collection of strips whose Minkowski sums with the tool cover the entire
region.  A \emph{minimum strip cover} is a strip cover with the fewest strips.

\begin{lemma} \label{strip LB}
The size of a minimum strip cover is a lower bound on the number of turns in a
cycle cover (or tour) of the region.
\end{lemma}

\begin{proof}
Any cycle cover induces a strip cover by extending each edge to have maximal
length.  The number of strips in this cover equals the number of turns in the
cycle cover.
\hfill\end{proof}

\begin{figure}[htb]
\leavevmode
\begin{center}
\leavevmode
\epsfig{file=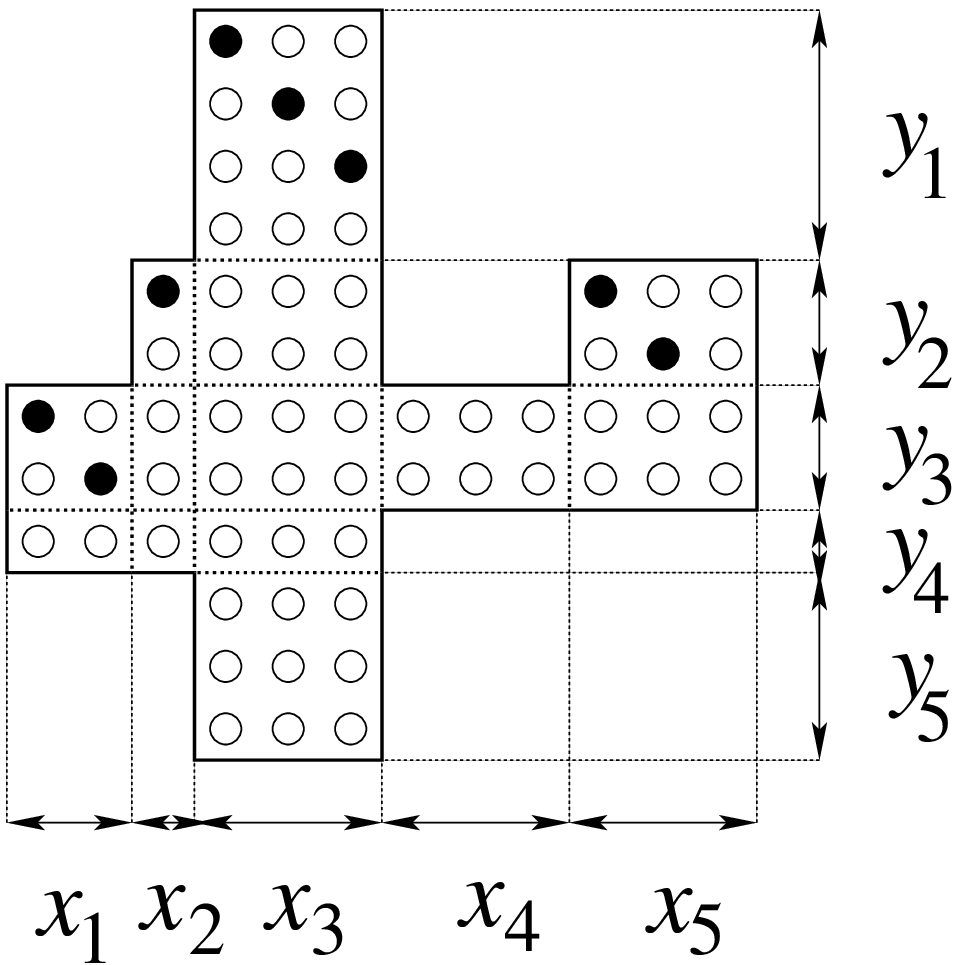,width=.40\linewidth} \hfill
\epsfig{file=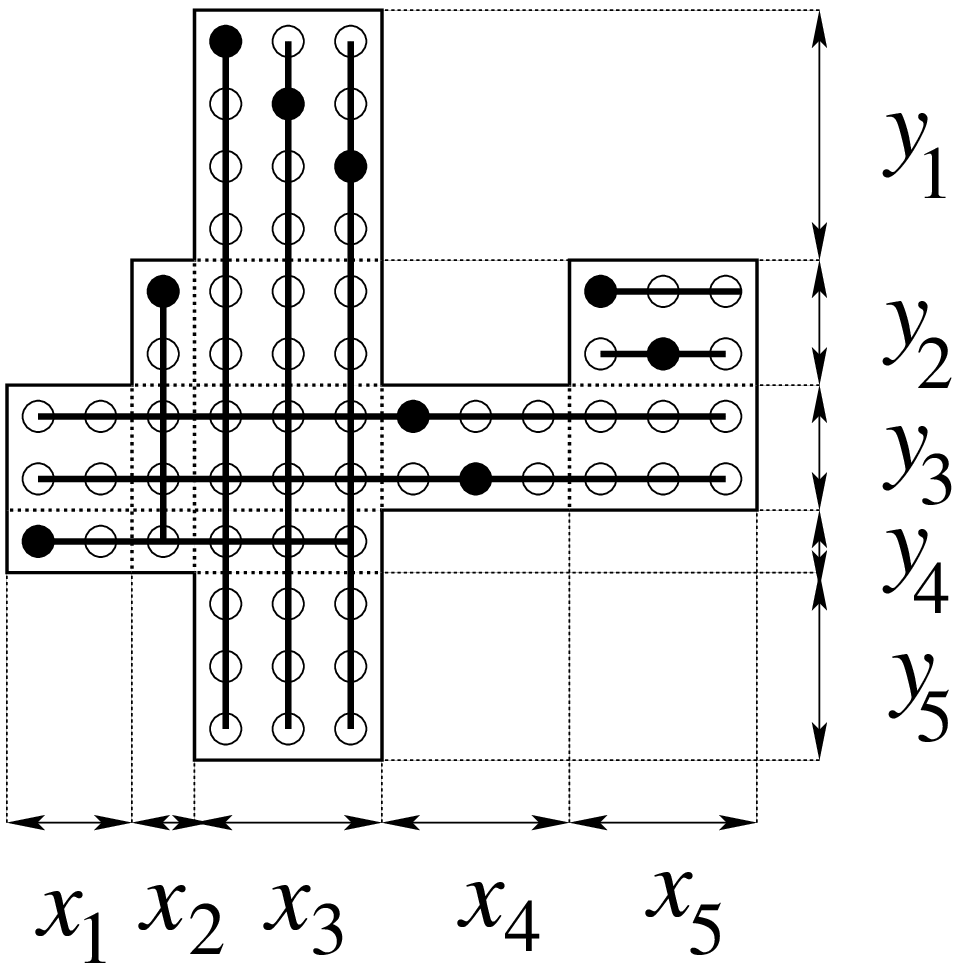,width=.40\linewidth} 
\end{center}
\caption{(Left) An orthogonal region, its subdivision
into axis-parallel strips, and a resulting greedy
rook cover (indicated by black pixels).
(Right) 
An optimal strip cover, and an 
optimal rook cover (indicated by black pixels).}
\label{fi:koenig}
\end{figure}

In the discrete milling problem, a related notion is a ``rook placement.''
A \emph{rook} is a marker placed on a pixel, 
which can \emph{attack} every pixel to which
it is connected via a straight axis-parallel path inside the region.  
A \emph{rook placement} is a collection of
rooks no two of which can attack each other.
See Figure~\ref{fi:koenig} for an illustration; this tool
will be used in Theorem~\ref{th:polynomial}, based on the following lemma.

\begin{lemma} \label{rook LB}
The size of a maximum rook placement is a lower bound on the number of turns
in a cycle cover (or tour) for discrete milling.
\end{lemma}

\begin{proof}
Consider a rook placement and a cycle cover of the region, which must in
particular cover every rook.  Suppose that one of the cycles visits rooks
$q_1, \dots, q_k$ in that order.  No two rooks can be connected by a single
straight axis-parallel line segment, so the cycle must turn inbetween 
each rook, for a total of at least $k$
turns.  Because each rook is traversed by at least one cycle, the number of
turns (and hence the number of segments in a tour) is at least the number 
of rooks.
\hfill\end{proof}

In the integral orthogonal milling problem, the notions of strip cover and
rook placement are dual and efficient to compute:

\begin{lemma} \label{strip cover integral orthogonal}
For integral orthogonal milling, a minimum strip cover and a maximum 
rook placement have equal size. For a polygonal region with $n$ edges
and $N$ pixels they can be computed in time $O(N^{2.376})$
or $O(n^{2.5}\log N)$.
\end{lemma}

\begin{proof}
For the case of $N\in O(n)$, the claim follows
from Proposition 2.2 in \cite{hm-aahos-91}: 
We rephrase the rook-placement problem as a matching problem in a
bipartite graph $G=(V_1,V_2,E)$.  Let the vertices in $V_1$ correspond to
vertical strips, and the vertices in $V_2$ correspond to horizontal
strips. An edge $e=(u,v)\in E$ exists if the vertical strip
corresponding to $u$ and the horizontal strip corresponding to $v$
have a pixel in common (i.e., the strips cross). It is easy to see
that a maximum-cardinality matching in this bipartite graph
corresponds to a rook placement: each edge $(u,v)$ in the matching
corresponds to the unique pixel that vertical strip $u$ and horizontal
strip $v$ have in common.

Similarly, observe that the minimum strip-cover problem is equivalent
to a minimum vertex-cover problem in the bipartite graph defined
above. Each strip in the strip cover defines a vertex in the vertex
cover. The requirement that each pixel must be covered by at least one
strip is equivalent to the requirement that each edge of the graph
must be covered by at least one vertex.

By the famous K\"onig-Egerv\'ary theorem, the maximum cardinality
matching in a bipartite graph is equal in size to the minimum vertex cover,
and therefore both can be solved in time polynomial in the size of the
graph; more precisely, this can be achieved in time $O(N^{\omega})$, 
the time needed
for multiplying two $N\times N$ matrices, for example $\omega=2.376$; 
see the paper \cite{im-dpamb-81},
or the survey in chapter 16 of \cite{s-co-03}, 
which also lists other, more elementary methods.

To get the claimed running time even for ``large''
$N$, using implicit encoding, we decompose the region into ``thick strips'' by
conceptually coalescing adjacent horizontal strips with the same horizontal
extent, and similarly for vertical strips.  In other words, thick strips are
bounded by two vertices of the region, and hence there are only 
$O(n)$ of them.  
We define the same bipartite graph but add a weight to each vertex
corresponding to the width of the strip (i.e., the number of strips coalesced).
Instead of a matching, in which each edge of the graph is either included in
the matching or not, we now have a multiplicity for each edge, 
which is the minimum
of the weights of its two endpoints.  The interpretation is that an edge
corresponds to a rectangle in the region (the intersection of two thick
strips), and the number of rooks that can be placed in such a rectangle is at
most the minimum of its width and height.

The weighted-matching problem we consider 
is that each edge can be included in the matching with a 
multiplicity up to its weight. 
Furthermore, the sum of the included 
multiplicities of edges incident to a
vertex cannot exceed the weight of the vertex.  A weighted version of the
K\"onig-Egerv\'ary theorem states that the minimum-weight vertex cover is equal to
the maximum-weight matching.  (This weighted version can be easily proved using
the max-flow min-cut theorem.)  Both problems can be solved in polynomial time
using a max-flow algorithm, on a modified graph in which a source vertex $s$ is
added with edges to all vertices in $V_1$, of capacity equal to the weight of
the vertex, and a sink vertex $t$ is added with edges to it from all vertices
in $V_2$ with capacity equal to the vertex capacity. Edges between $V_1$ and
$V_2$ are directed from $V_1$ and have capacity equal to the weight of the
edge. Currently, the best known running time is 
$O(\sqrt{n}m\log nW)$ for a bipartite graph with $n$ vertices,
$m$ edges, and maximum weight $W$ \cite{gt-fsanp-89,s-co-03}.
For our purposes, this yields a complexity of $O(n^{2.5}\log N)$.
\hfill\end{proof}

Note that using weights on the edges is crucial for the correctness
of our objective; moreover, this has a marked effect on the complexity
of the problem: Finding a minimum number of axis-parallel rectangles
(regardless of their size) that covers an integral orthogonal polygon 
is known to be an NP-complete problem, even for the case
of polygon without holes \cite{cr-occop-89a}.

For general discrete milling, it is possible to approximate an optimal strip
cover as follows.  Greedily place rooks until no more can be placed 
(i.e., until there is no unattackable vertex).  This means that every
vertex is attackable by some rook, so by replacing each rook with all
possible strips through that vertex, we obtain a strip cover of size $\avgdir$
times the number of rooks, where $\avgdir$ is the maximum
degree of the underlying graph.  (We call this type of strip cover a \emph{star
cover}.)  But each strip in a minimum strip cover can only cover a single
rook, so this is a $\avgdir$-approximation to the minimum strip cover.  We
have thus proved 

\begin{lemma} \label{star cover}
In discrete milling, the number of stars in a greedy star cover is a lower
bound on the number of strips, and hence serves as a $\avgdir$-approximation
algorithm for minimum strip covers.
Computing a greedy star cover can be done in time $O(N)$. 
\end{lemma}

\begin{proof}
Loop over the vertices of the underlying graph.  Whenever an unmarked
vertex is found, add it to the list of rooks, and mark it and all vertices
attackable by it.
Now convert each rook into a star as in the proof of Lemma \ref{rook LB}.
Each edge is traversed only once during this process.
\hfill\end{proof}

\section{Approximation Algorithms}

We employ four main approaches to building approximation algorithms, repeatedly
in several settings:

  \begin{enumerate}
  \item[(i)] \textbf{Star cover + doubling + merging}

        The simplest but most generally applicable idea is to cover the
        region by a collection of stars.  ``Doubling'' these stars results
        in a collection of cycles, which can then be merged into a tour
        using general techniques.

  \item[(ii)] \textbf{Strip cover + doubling + merging}

        Tighter bounds can be achieved by covering directly with strips
        instead of stars.  Similar doubling and merging steps follow.

  \item[(iii)] \textbf{Strip cover + perfect matching of endpoints + merging}

        The covering of the region is done by the strip cover.
        To connect these strips into cycles, we find a minimum-weight
        perfect matching on their endpoints.  This results in a cycle cover,
        which can be merged into a tour using general techniques.

  \item[(iv)] \textbf{Boundary tour + strip cover + perfect matching of odd-degree vertices}

        Again coverage is by a strip cover, but the connection is done
        differently.  We add a tour of the boundary (by merging an optimal
        boundary cycle cover), and attach each strip to this tour on each end.
        The resulting graph has several degree-three vertices, which we fix
        by adding a minimum-weight matching on these vertices.
  \end{enumerate}

\subsection{Discrete Milling}
\label{sub:disc}

As described in the Preliminaries, we consider two scenarios:
While general discrete milling focuses on vertices (and thus resembles
the TSP), thin discrete
milling requires traveling a set of edges, making it similar to the CPP.

\subsubsection{General Discrete Milling}
\label{sub:disc.gen}
Our most general approximation algorithm for the discrete milling problem 
runs in linear time.  First we take a star cover
according to Lemma \ref{star cover}, which approximates an optimal strip cover
to within a factor of $\avgdir$.  
Then we tour the stars using an efficient method
described below.  Finally we merge these tours using Theorem \ref{merging
cycles}.

%\comment{REFEREE 1: The approx result for discrete milling depends on
  %avg deg, avg number of directions, etc in the graph; Can the authors
  %give some idea about the range of values these quantities can
  %assume, so that one can judge better the quality of the result?}
%DONE. Added a remark in the introduction.

\begin{figure}[htb]
\leavevmode

\begin{center}
\leavevmode
\epsfig{file=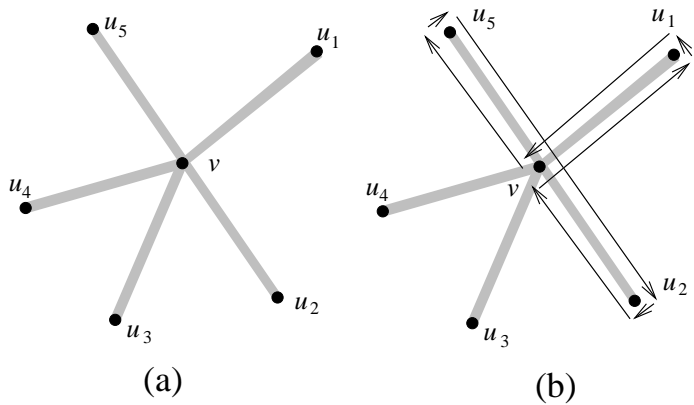,width=.7\linewidth} 
\end{center}

\caption{(a) A star of degree five around vertex $v$.
$(v,u_1)$, $(v,u_3)$, $(v,u_4)$ are half strips, $(u_2,u_5)$
is a full strip.
(b) A covering with three edges for a half strip, and a covering
with five edges for the full strip.}
\label{fi:star}
\end{figure}

We tour each star emanating from a vertex $v$ using the following
method---see Figure~\ref{fi:star}.
Consider a strip $s$ in
the star, and suppose its ends are the vertices $u_i$ and $u_j$.  A strip
having both of its endpoints distinct from $v$ is called a {\em full
strip}; a strip one of whose endpoints is equal to $v$ is
called a {\em half strip}.  Half strips are covered by three edges,
$(v,u_j)$, $(u_j,u_j)$, and $(u_j,v)$, making a U-turn at endpoint $u_j$.  
(This covering is shown for the half 
strip $(v,u_1)$ in Figure~\ref{fi:star}(b).)
Full strips are covered by five edges, $(v,u_i)$, $(u_i,u_i)$, 
$(u_i,u_j)$, $(u_j,u_j)$,
and $(u_j,v)$, with U-turns at both ends, $u_i$ and $u_j$.  
(This covering is shown for the full strip $(u_5,u_2)$ in 
Figure~\ref{fi:star}(b).) Now we have
several paths of edges starting and ending at $v$.  By joining their
ends we can easily merge these paths into a cycle.

The number of turns in this cycle is 3 times the number of half
strips, plus 5 times the number of full strips.  This is equivalent to
the number of distinct directions at $v$ plus 2 times the degree of
$v$.  (The number of directions at $v$ is equal to the number of full
strips plus the number of half strips, by definition.  The degree of
$v$ is equal to 2 times the number of full strips plus the number of
half strips.)  Lemma \ref{star cover} implies that the number of stars
is a lower bound on the number of turns in a cycle cover of the
region, proving the following:

\begin{theorem} \label{discrete cycle cover approx}
There is an $O(N)$-time $(2\maxdeg+\avgdir)$-approximation for finding
a minimum-turn cycle cover in discrete milling.  Furthermore, the 
maximum coverage
of a vertex (i.e., the maximum number of times a vertex is swept) 
is $\maxdeg$, and the cycle cover is a $\maxdeg$-approximation on
length. 
\end{theorem}

\begin{proof}
As the star cover, by definition, contains all vertices, the cycle
cover obtained by traversing the stars also does.  As stated above,
the number of turns in each cycle covering a star is the number of
directions at $v$, plus 2 times the degree of $v$.  Summing over all
stars, we get the claimed approximation bound.  The running time
follows directly from Lemma \ref{star cover}. Deriving the values for maximum
coverage and overall length is straightforward.
\hfill\end{proof}

\begin{corollary} \label{discrete tour approx}
There is a linear-time $(2\maxdeg + \avgdir + 2)$-approximation for
minimum-turn discrete milling.  Furthermore, the maximum coverage of a vertex
is $\maxdeg$, and the tour is a $\maxdeg$-approximation on length.
\end{corollary}

\begin{proof}
We apply Theorem \ref{merging cycles}.  The number of cycles to be
merged is the number of stars, which by Lemma \ref{star cover} is a
lower bound on the number of turns in a tour of the region.  We pay at
most two turns per cycle for the merge.  There is no additional cost
of length due to the merge, as the stars form a connected graph.
\hfill\end{proof}

%Note that, in particular, these algorithms give a $6$-approximation on length
%if the discrete milling problem comes from a planar graph, because the average
%degree of a planar graph is bounded by 6.

\subsubsection{Thin Discrete Milling}
\label{sub:disc.thin}
As described in Section~\ref{prelim}, a more special structure
arises if the structure to be milled consists of a connected set of
``channels'' that have to be milled. In this case, achieving a strip cover
is trivial:

\begin{lemma} \label{strip CPP}
In thin discrete milling a strip cover can be obtained
in linear time by merging edges that are collinear at some vertex.
\end{lemma}

Using method (ii) described at the beginning of the section,
we get the following approximation results:

\begin{theorem} \label{thin.CPP}
There is a $4$-approximation of 
complexity $O(n\log n)$ 
for computing a minimum-turn cycle cover for a graph with $n$ edges,
and a $6$-approximation 
of the same complexity for computing minimum-turn tours.  
\end{theorem} 

\begin{proof}
Clearly, the number of strips is a lower bound on the cost of any cycle
cover or tour.
Turning each strip into a cycle with two u-turns, i.e., 4 turns,
yields a cycle cover
within a factor 4 of the optimum. Merging these cycles at a cost of
2 turns per merge yields a tour within a factor of 6 times the optimum,
as each cycle has 4 turns. 
\hfill\end{proof}

Clearly, all edges get covered twice
(yielding a bound of 2 on the simulataneous length approximation)
and no vertex gets covered more than $2\avgdir$ times.

Using the more time-consuming method (iii),
we get better approximation factors for the turn cost:

\begin{theorem} \label{thin.CPP.iii}
There is a $1.5$-approximation of 
complexity $O(n^3)$ 
for computing a minimum-turn cycle cover for a graph with $n$ edges,
and a $3.5$-approximation 
of the same complexity for computing minimum-turn tours.  
\end{theorem} 

\begin{proof}
As before, we can compute an optimal strip cover $S$ in linear time.
Analogous to the approach in Section~\ref{ssec:boundary}
define a weight function between endpoints of strips, taking
into account the direction when leaving a strip.
Clearly, any feasible tour consists of two different matchings $M_1$
and $M_2$ between strip endpoints; moreover, if $d(M_1)$
and $d(M_2)$ are the total weights of the edges in the matchings,
we get $d(M_1)+d(M_2)\leq d(T)$. It follows that for an optimal
matching $M$, we have $d(M)\leq$ {\sc opt}$/2$.
By construction, the edges of $M$ and the strips of $S$
induce a 2-factor of the vertices that covers all edges.
Thus, we get a cycle cover of cost at most $1.5$ {\sc opt}. 

Now consider the $c$ cycles in a cycle cover. $c$ is at most the number
of strips in $S$, which is a lower bound on the cost of an optimal tour.  
As the cost of merging the $c$ cycles is at most $2c-2$, we get a
a total cost of not more than $3.5$ {\sc opt}.
\hfill\end{proof}

\subsection{Integral Orthogonal}
\label{sub:int}

%\comment{REFEREE 2: In contrast to the rest of the paper, this
        %section had an unusually large number minor English errors.  A
        %careful review might be a good idea.}
As mentioned in the preliminaries, just the pixel count $N$ may not be
a satisfactory measure for the complexity of an algorithm, as the original
region may be encoded more efficiently by its boundary, and a tour
may be encoded by structuring it into a small number of
pieces that have a short description. It is possible to use the above
ideas for approximation algorithms in this extended
framework. We describe how this can be done for
the integral orthogonal case, where the set of pixels is bounded
by $n$ boundary edges.

\begin{theorem}
\label{th:polynomial}
There is a $10$-approximation of (strongly polynomial)
complexity $O(n\log n)$ 
for computing a minimum-turn cycle cover for a region of pixels bounded by
$n$ integral axis-parallel segments, and a 
$12$-approximation of the same complexity for computing minimum-turn tours.  
In both cases, the maximum
coverage of a point is at most $4$, so the algorithms are also
$4$-approximations on length.

For the special case in which the boundary is connected
(meaning that the region has no holes), the complexities
drop to $O(n)$.
\end{theorem}

\begin{proof}
The basic idea is to find a greedy rook cover, 
then use it to build an approximate tour.
Lemma~\ref{star cover} still holds,
and each strip in a star (as described in the previous section)
will be a full strip. The 
approximation ratios follow as special cases of 
Theorem~\ref{discrete cycle cover approx}: 
In this case, $\avgdir = 2$ and $\maxdeg = 4$.
It remains to show how we can find a greedy rook cover in
the claimed time. 

%\comment{REFEREE 2: The placement of figure 4.4 is odd, given that it
        %does not seem to be referenced in Section 4, but is important here.
        %Perhaps a simplified version of the figure could be placed in
        %Section 4, and the original moved here??}
%xxx Fixed by referencing in text.

Refer back to Figure~\ref{fi:koenig}.
Subdivide the region by the $n$ vertical chords through
its $n$ vertices, resulting into at most $n$ vertical strips
$X_1,\ldots,X_n$, of widths $x_1,\ldots,x_n$. Similarly, 
consider a subdivision by the $n$ horizontal chords through the
$n$ vertices into at most $n$ horizontal strips $Y_1,\ldots,Y_n$,
of width $y_1,\ldots,y_n$. In total, we get a subdivision into
at most $n^2$ cells $C_{ij}$. Despite of this quadratic
number of cells, we can deal with the overall problem in
near-linear time:
Note that both subdivisions can be found in time $O(n\log n)$.
For the case of a connected boundary, Chazelle's linear-time
triangulation algorithm \cite{c-tsplt-91a} implies a complexity of $O(n)$.

Choose any cell $C_{ij}$, which is a rectangle of size
$x_i\times y_j$. Then $r_{ij}=\min\{x_i,y_j\}$
rooks can be placed greedily along the diagonal of 
$C_{ij}$, without causing any interference; such a set 
of rooks can be encoded as one ``fat'' rook, described
by its leftmost uppermost corner $(\xi_{ij},\eta_{ij})$, and 
its width $r_{ij}$. Then 
the strip $X_i$ can contain at most $x_i-r_{ij}$ additional
rooks, and $Y_j$ can contain at most $y_j-r_{ij}$ rooks. Therefore,
replace $x_i$ by $x_i-r_{ij}$, and $y_j$ by $y_j-r_{ij}$.
This changes the width of at least one of the strips to zero,
effectively removing it from the set of strips.
After at most $2n-1$ steps of this type, all horizontal
or all vertical strips have been removed, implying that
we have a maximal greedy rook cover.
%%\comment{REFEREE 2: By the way, note that this "greedy" method is not very greedy,
        %because the cells are chosen in arbitrary order.}
%xxx Right; but I didn't change anything, as this doesn't ask for anything.

It is straightforward to see that for a fat rook at position
$(\xi_{ij},\eta_{ij})$ and width $r_{ij}$, there is a canonical
set of $r_{ij}$ cycles with $10$ edges each
that covers every pixel that can be attacked
from this rook. Furthermore, there is a ``fat''
cycle with at most $12r_{ij}-2$ turns that is obtained
by a canonical merging of the $r_{ij}$ small cycles.
Finally, it is straightforward to merge the fat cycles.
\hfill\end{proof}

If we are willing to invest more
time for computation, we can find an optimal rook cover
(instead of a greedy one). As discussed in the proof
of Lemma~\ref{strip cover integral orthogonal},
this optimal rook cover yields an optimal strip cover. 
An optimal strip cover can be used to get a $6$-approximation, 
and the new running time is $O(n^{2.5}\log n)$
or $O(N^{2.376})$.

\begin{theorem}
\label{thm:6-approx}
There is an $O(n^{2.5}\log N)$-time or $O(N^{2.376})$-time
algorithm that computes a milling tour with
number of turns within $6$ times the optimal, and with length within
$4$ times the optimal.
\end{theorem}

\begin{proof}
Apply Lemma \ref{strip cover integral orthogonal} to find an optimal
strip cover of the region. (See Figure~\ref{fi:koenig}.)
As described in the proof of that lemma, the cardinality of
an optimal strip cover is equal to the cardinality of an
optimal rook cover.
As stated, the number of strips is a lower bound on the number
of turns in a cycle cover or tour.

Now any strip from $u$ to $w$ is covered by a ``doubling'' cycle 
with edges $(u,w)$, $(w,w)$, $(w,u)$, $(u,u)$.
This gives a $4$-approximation to minimum-turn cycle covers.  
Finally apply Corollary \ref{cycle-cover-merging} to 
get a $6$-approximation to minimum-turn tours. 

The claim about coverage (and hence overall length) follows
from the fact that an optimal strip cover has maximum coverage 2,
hence the cycle cover has maximum coverage 4.
\hfill\end{proof}

By more sophisticated merging procedures, it is possible to 
reduce the approximation factor for tours to a figure closer to $4$.
Note that
in the case of $N$ being large compared to $n$, 
the above proof grossly overestimates the 
cost of merging,
as all cycles within a fat strip allow merging at no additional cost.  
However,  our best approximation algorithm achieves a factor 
less than $4$ and uses a different strategy.  

\begin{theorem} \label{3.75-approx integral orthogonal}
For an integral orthogonal polygon with $n$ edges and $N$ pixels,
there are $2.5$-approximation algorithms, with running times
$O(N^{2.376}+n^3)$ and $O(n^{2.5}\log N + n^3)$, 
for minimum-turn
cycle cover, and hence there is a polynomial-time $3.75$-approximation 
for minimum-turn tours.
\end{theorem}

\begin{proof}
As described in Lemma~\ref{strip cover integral orthogonal},
find an optimal strip cover $S$, in time
$O(N^{2.376})$
or $O(n^{2.5}\log N)$. Let $s$ be its cardinality
and let {\sc opt} be the cost of an optimal tour,
then {\sc opt}$\geq s$.

Now consider the end vertices of the strip cover. By construction,
they are part of the boundary. Because 
any feasible tour $T$ must encounter each pixel, 
and cannot cross the boundary,
any endpoint of a strip is either crossed orthogonally, or the 
tour turns at the boundary segment. In any case, a tour must
have an edge that crosses an end vertex orthogonally to the strip.
(Note that this edge has zero length in case of a U-turn.)

As in Section~\ref{ssec:boundary} and the proof of Theorem~\ref{thin.CPP.iii},
define a weight function between endpoints of strips, taking
into account the direction when leaving a strip.
Again any feasible tour consists of two different matchings $M_1$
and $M_2$ between strip endpoints, and for an optimal
matching $M$, we have $d(M)\leq$ {\sc opt}$/2$.

Computing such a matching can be achieved as follows.
Note that for $N$ pixels, an optimal strip cover has
$O(\min\{\sqrt{N},n\})$ strips; by matching endpoints of neighboring strips
within the same fat strip, we are left with $O(n)$ endpoints.
As described in the proof of Lemma~\ref{strip cover integral orthogonal},
the overall cost for computing the link distance
between all pairs of endpoints can be achieved in $O(\min\{N\log N,n^2\log n\})$. Computing a minimum-weight perfect matching
can be achieved in time $O(\max\{N^{1.5},n^3\})$.

The edges of $M$ and the strips of $S$ induce 
a 2-factor of the endpoints. Because any matching edges leaves
a strip orthogonally, we get at most 2 additional turns at each strip
for turning each 2-factor into a cycle. The total number of turns
is $2s+w(M)\leq 2.5\cdot${\sc opt}. Because the strips cover the
whole region, we get a feasible cycle cover.

Finally, we can use Corollary~\ref{cycle-cover-merging} to turn 
the cycle cover into a tour. By the corollary, this tour does 
not have more than $3.75\cdot${\sc opt} turns.
\hfill\end{proof}

The class of examples in Example~\ref{ex:donut} shows
that the cycle cover algorithm may use $2\cdot${\sc opt}
turns, and the tour algorithm may use $3\cdot${\sc opt} turns, 
assuming that no special algorithms are used for matching
and merging. Moreover, the same example shows
that this $3.75$-approximation algorithm does not 
give an immediate  length bound on the resulting tour:  

\begin{exam}\label{ex:donut}
The class of regions shown in Figure~\ref{fi:bad}
may yield a heuristic cycle cover with $2\cdot${\sc opt} turns,
and a heuristic tour with $3\cdot${\sc opt} turns.
\end{exam}

The region consists of 
a ``square donut'' of width $k$. An optimal strip cover
consists of $4k$ strips; an optimal matching
of strip ends yields a total of $8k+2$ turns, and we get a total
of $2k$ cycles. 
(In Figure~\ref{fi:bad}(a), only the vertical strips and their
matching edges are shown to keep the drawing cleaner.)
If the merging of these cycles is done badly (by merging cycles
at crossings and not at parallel edges),
it may cost another $4k-2$ turns, for a total of $12k$ turns.
As can be seen from Figure~\ref{fi:bad}(b), there is a feasible
tour that uses only $4k+2$ turns. This shows that optimal tours
may have almost all turns strictly inside of the region.
Moreover, the same example shows
that this $3.75$-approximation algorithm does not 
give an immediate  length bound on the resulting tour.
However, we can use a local modification argument to show the following:

\begin{figure}[htb]
\leavevmode
\begin{center}
\leavevmode
\epsfig{file=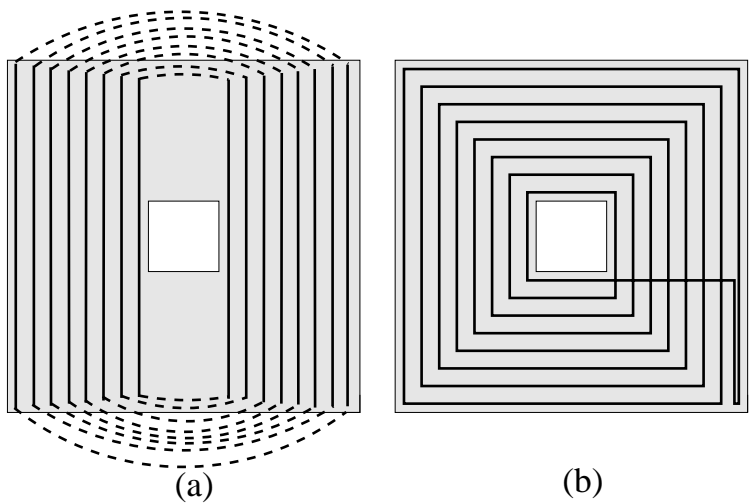,width=.78\linewidth} 
\caption{A bad example for the 3.75-approximation algorithm:
(a) Half the cycles constructed by the algorithm.
(b) An optimal tour.
}
\label{fi:bad}
\end{center}
\end{figure}

\begin{theorem} \label{th:brackets}
For any given feasible tour (or cycle cover)
of an integral orthogonal region, there
is a feasible tour (or cycle cover)
of equal turn number that covers 
each pixel at most four 
times. This implies a performance ratio of 4 on the total length.
\end{theorem}

\begin{figure}[htbp]
\leavevmode
\begin{center}
\leavevmode
\epsfig{file=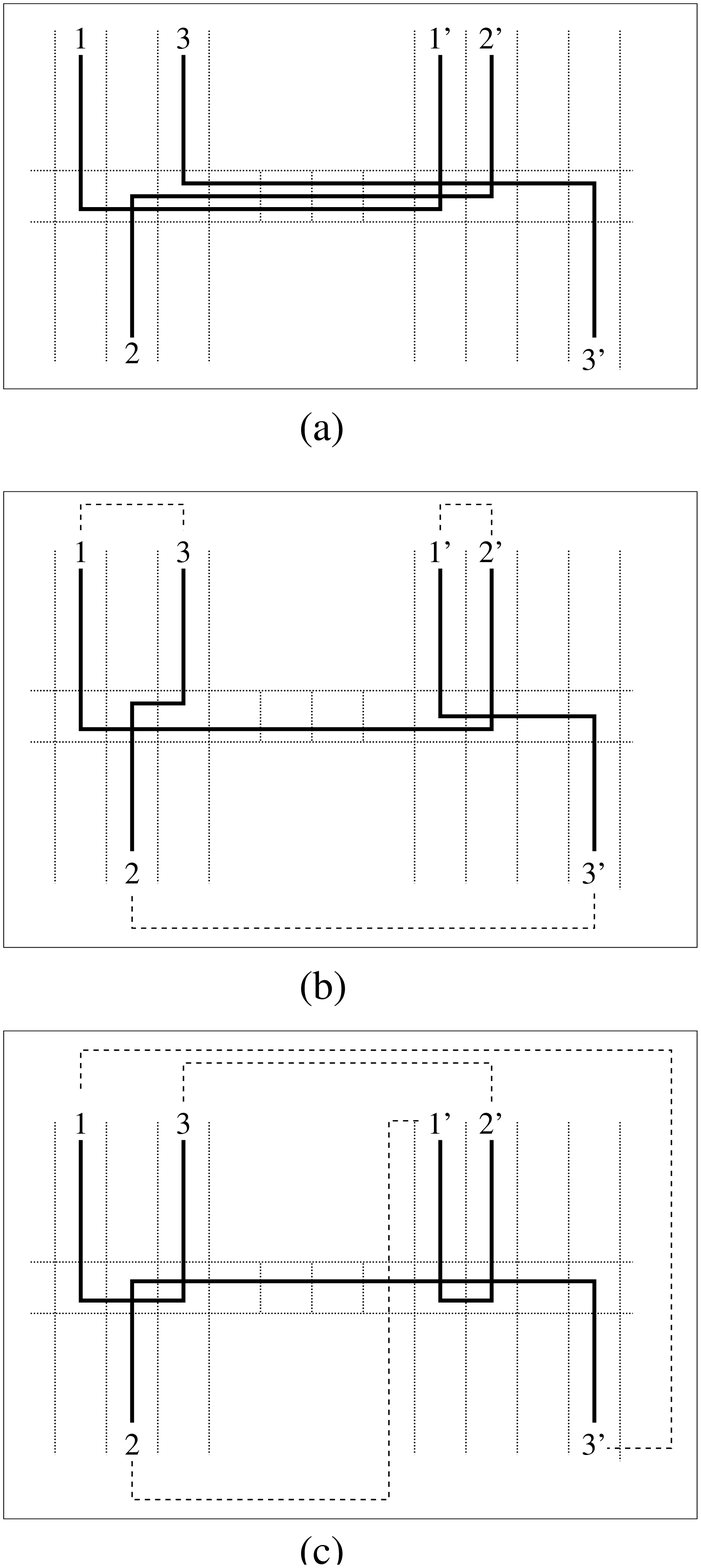,width=.5\linewidth} 
\caption{Rearranging a tour to ensure that no pixel is covered
more than three times in each direction:
(a) A set of horizontal edges that covers some pixel three times.
(b) A rearranged tour, if the second matching between endpoints
connects two left and two right endpoints.
(c) A rearranged tour, if the second matching between endpoints
connects any left with a right endpoint.
}
\label{fi:brackets}
\end{center}
\end{figure}

\begin{proof}
See Figure~\ref{fi:brackets}.
Suppose there is a pixel that is covered at least five times.
Then there is a direction (say, horizontal) in which it is covered
at least three times. Let there be three horizontal segments
$(1,1')$, $(2,2')$, $(3,3')$ 
covering the same pixel, as shown in Figure~\ref{fi:brackets}(a);
we denote by $1$, $2$, $3$ the endpoints to the left of the pixel, and by
$1'$, $2'$, $3'$ the endpoints to the right of the pixel.

Now consider the connections of these points by the rest of the tour,
i.e., a second matching between the points $1$, $2$, $3$, $1'$, $2'$, $3'$ that forms
a cycle, when merged with the first matching 
$(1,1')$, $(2,2')$, $(3,3')$. 
This second matching is shown dashed in Figure~\ref{fi:brackets}(b,c).
We consider two cases, depending on the structure of the second matching.

In the first case, there are two right endpoints that are matched, say $1'$ and $2'$.
Then there must be two left endpoints that are matched; because both matchings
must form one large cycle, these cannot be $1$ and $2$. Without loss of generality,
we may assume they are $1$ and $3$. Thus, $2$ and $3'$ must be matched, as shown
in Figure~\ref{fi:brackets}(b). Then we can replace 
$(1,1')$, $(2,2')$, $(3,3')$
by
$(1,2')$, $(2,3)$, $(1',3')$, respectively, 
which yields a feasible tour with the same number of turns,
but with some pixels being covered fewer times, and no pixel
being covered more times than was the case in the original tour.

In the other case, all right endpoints are matched with left endpoints.
Clearly, $1'$ cannot be matched with 1; without loss of generality,
we assume it is matched with 2, as shown in Figure~\ref{fi:brackets}(c).
Then the cycle condition implies that the second matching is
$(1,3')$, $(2,1')$, $(3,2')$. This allows us to replace 
$(1,1')$, $(2,2')$, $(3,3')$
by
$(1,3)$, $(2,3')$, $(1',2')$, respectively, 
again producing a feasible tour with the same number of turns,
but with some pixels being covered fewer times, and no pixel
being covered more times than was the case in the original tour.

This can be repeated until no pixel is covered more than four times.
As the above procedure can be carried out as part of the merging
phase (i.e., after an optimal weighted matching has been found),
the overall complexity is not affected. Furthermore, it is straightforward
to see that it also works for the case of ``thick'' strips,
where $N$ is large compared to $n$, by treating parallel edges in a
thick strip simultaneously.
\hfill\end{proof}

\subsection{Nonintegral Orthogonal Polygons}
\label{sub:nonint}

Nonintegral orthogonal polygons present a difficulty in that 
no polynomial-time algorithm is known
to compute a minimum strip cover for such polygons.
Fortunately, however, we can use the boundary tours from 
Section \ref{ssec:boundary} to the approximation factor of $12$ 
from Theorem \ref{th:polynomial} 
for the integral orthogonal case.  

\begin{theorem}\label{th:fractional}
In nonintegral orthogonal milling of a polygonal region
with $n$ edges and $N$ pixels, there is a polynomial-time
$4.5$-approximation for mi\-ni\-mum-turn cycle covers and $6.25$-approximation 
for
mi\-ni\-mum-turn tours, with a simultaneous performance guarantee
of 8 on length and cover number. The running time is $O(N^{2.376}+n^3)$,
or $O(n^{2.5}\log N+n^3)$.
\end{theorem}

\begin{proof}
Take the $2.5$-approximate cycle cover of the integral pixels
in the region as in Theorem \ref{3.75-approx
integral orthogonal}; for a tour, turn it into a $3.75$-approximate tour.
This may leave a fractional portion along the boundary uncovered.
See Figure~\ref{fi:fractional}.

\begin{figure}[htb]
\leavevmode
\begin{center}
\leavevmode
\epsfig{file=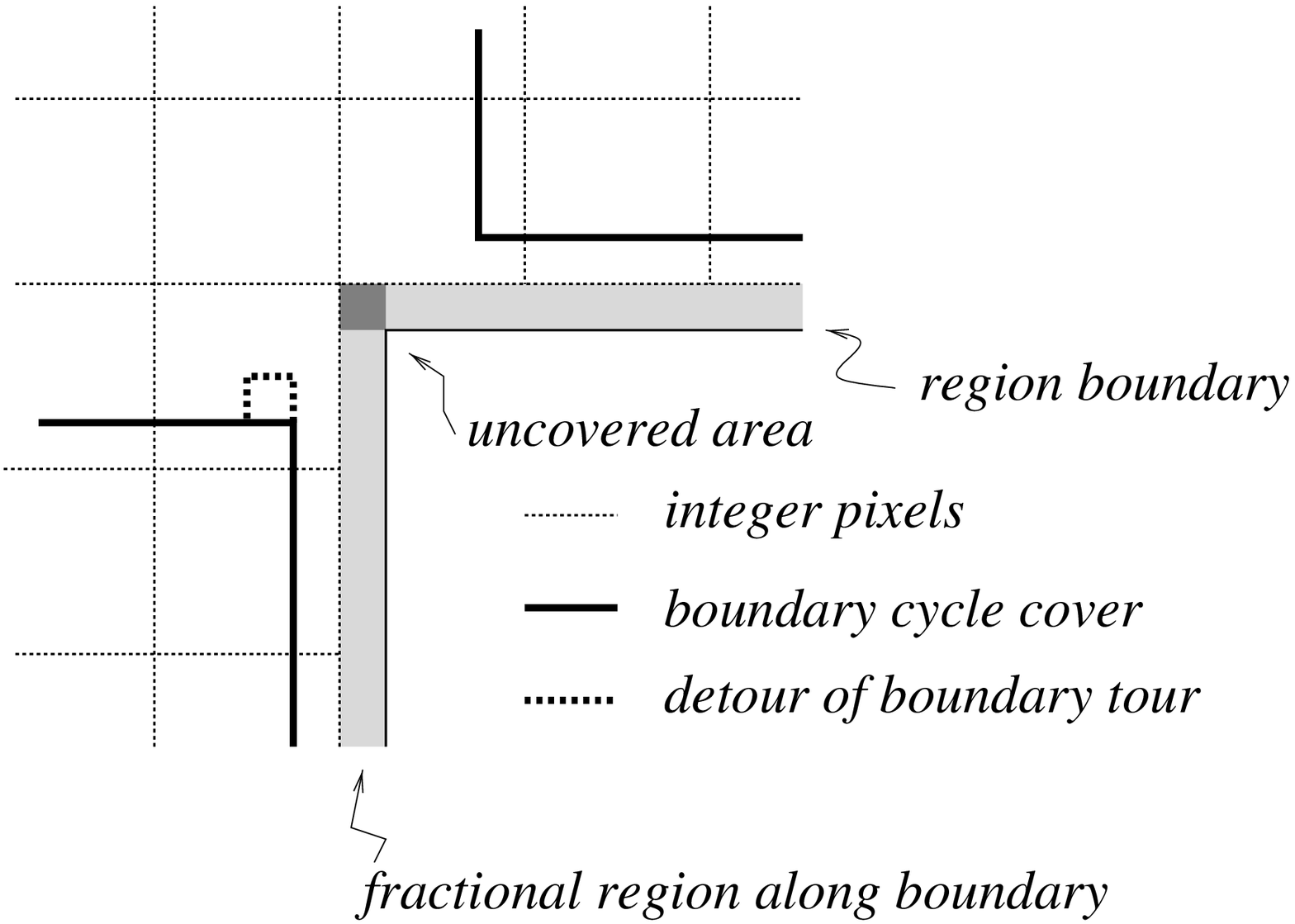,width=0.58 \linewidth}
\caption{Milling a nonintegral orthogonal polygon}
\label{fi:fractional}
\end{center}
\end{figure}

Now add in an optimal cycle cover of the boundary which comes from 
Theorem \ref{th:strong}.
This may only leave fractional boundary pieces uncovered
that are near reflex vertices of the boundary,
as shown in Figure~\ref{fi:fractional}. Whenever this  happens,
there must be a turn of the boundary cycle cover on both side
of the reflex vertex. The fractional patch can be covered at the
cost of an extra two turns, which are charged to the two turns
in the boundary cycles. Therefore, the modified boundary cover
has a cost of at most $2\cdot${\sc opt}. 
Compared to an optimal cycle cover of length
{\sc opt}, we get a cycle cover of length at most $4.5\cdot${\sc opt},
as claimed. For an optimal tour of length {\sc opt},
merging all modified boundary cycles into one cycle can be done
at a cost of at most 2 turns per unmodified boundary cycle, 
i.e., for a total of $\frac{1}{2}\cdot${\sc opt}.

Finally, the remaining two cycles can be merged at a cost of 2 turns.
This yields an overall approximation factor of $3.75+2.5=6.25$.
The claim on the cover number (and thus length) follows from applying
Theorem~\ref{th:brackets} to each of the two cycles.

The running times follow from Theorems \ref{th:strong}
and  \ref{3.75-approx integral orthogonal}.
\hfill\end{proof}

\subsection{Milling Thin Orthogonal Polygons}
\label{sub:thin}
%\comment{SANDOR: I rewrote Section 5.4. Please proofread!}
In this section we consider the special case of milling {\em thin\/}
polygons.  Again, we focus on the integral orthogonal case.  Formally,
a thin polygon is one in which no axis-aligned 2$\times$2 square fits,
implying that each pixel has all four of its corners on the boundary
of the polygon. 
Intuitively, a polygon is thin if it is composed of a
network of width-1 corridors, where each pixel is adjacent to some
part of the boundary of the region, making this related to discrete milling.

\subsubsection{Basics of Thin Orthogonal Polygons}
\begin{figure}[htb]
\leavevmode
\begin{center}
\leavevmode
\epsfig{file=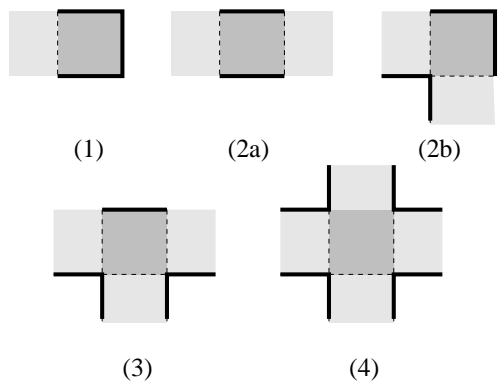,width=0.5 \linewidth}
\caption{Pixel types in a thin polygon.}
\label{fig:degree}
\end{center}
\end{figure}

Any pixel in the polygon has one, two, three or four neighbor 
pixels; we denote this number of neighbors as the degree of a pixel.
See Figure~\ref{fig:degree}.
Degree one pixels (1) are ``dead ends'', where the cutter has to make 
a u-turn. There are two types of degree two pixels,
without forcing a turn (2a) or with forcing a turn (2b); in either case,
applying Lemma~\ref{le:local} in an appropriate manner will suggest
that they should be visited in a canonical way:
after one neighbor, and before the other.
Neighbors of degree three pixels (3) form ``T'' intersections that 
force duplication of paths. Degree four pixels (4) are the only 
pixels in thin polygons that are not boundary pixels as 
defined in Section~\ref{sec:tools}; however, in the absence of 2$\times$2
squares of pixels, all their neighbors are of degree one or two.

In the following, we will use the ideas developed for boundary cycle
covers in Section~\ref{ssec:boundary} to obtain cycle covers for
thin polygons. The following is a straightforward 
consequence of Theorem~\ref{th:strong} and
Theorem \ref{merging cycles}. 

\begin{corollary}
\label{co:thin}
In thin orthogonal milling, there is an $O(n^3)$ algorithm for computing a
minimum-turn cycle cover, and an $O(n^3)$ 
1.5-approximation for computing a minimum-turn tour.
\end{corollary}

\begin{proof}
Apply the strongly polynomial algorithm described in Theorem~\ref{th:strong}
for computing a minimum cost boundary cycle cover. By definition,
this covers all pixels of degree one, two, and three. Moreover, 
degree four pixels are surrounded by pixels of degree one or two, 
implying that they are automatically covered as neighbors of those pixels,
when applying Lemma~\ref{le:local}. Using Theorem~\ref{merging cycles},
we can turn this into a tour, yielding the claimed approximation factor.
\hfill\end{proof}

More interesting is that we can do much better than 
general merging in the case of thin orthogonal milling. The idea is 
to decompose the induced graph into a number of cheap cycles,
and a number of paths.

\subsubsection{Milling Thin Eulerian Orthogonal Polygons}

We first solve the special case
of milling Eulerian polygons, that is, polygons that can be milled
without retracing edges of the tour, so that each edge in the induced
graph is traversed by the cutting tool
exactly once.  In an Eulerian polygon, all pixels have either
two or four neighbors, meaning there are no odd-degree pixels. 

Although one might expect that the optimal milling is one of the 
possible Eulerian tours of the graph, in fact, this is not always true, 
as Example~\ref{ex:thnsquares} 
points out.

\begin{exam}\label{ex:thnsquares}
There exist thin grid graphs, such that no turn-minimal tour of the graph 
is an Eulerian tour.  
\end{exam}

\begin{figure}[htb]
\leavevmode

\begin{center}
\leavevmode
\epsfig{file=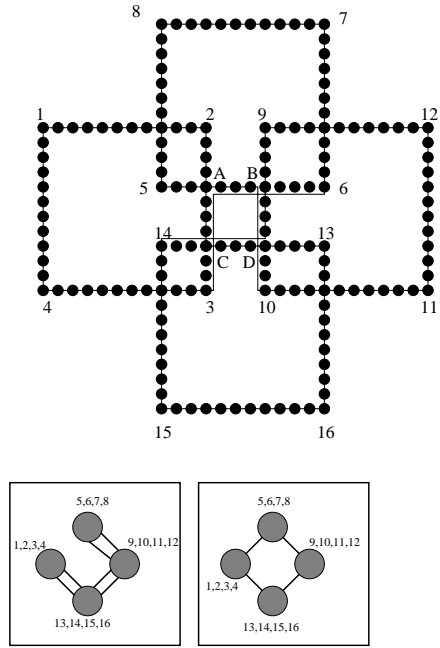,width=.44\linewidth} 
\end{center}

\caption{A thin Eulerian polygon consisting 
of four overlapping cycles (above). Shown symbolically below is how to 
obtain an overall tour by merging the four canonical cycles:
A tour obtained by iteratively merging cycles incurs a total of
16+6=22 turns (bottom left). An optimal tour has 16+4=20 turns 
(bottom right).  }
\label{fig:thinsquares}
\end{figure}

\begin{proof}
See Figure \ref{fig:thinsquares}.
Observe that an optimal milling is {\em not\/} an Eulerian Tour.
The best Eulerian Tour for this figure requires 22 turns,
as shown symbolically in the bottom left of the figure:  
Each cycle uses 4 turns and an additional 6 turns 
can be used to connect the four cycles together.
On the other hand, the optimal milling traverses the edges 
in the internal pixel twice, both times in the same direction:  
The order of turns is $1,2,C,13,16,15,14,D,9,12,11,10,B,5,8,7,6,A,3,4,1$,
and the structure is shown symbolically in the bottom right.
Thus, the optimal milling only requires 20 turns, where
each cycle uses 4 turns and an additional 4 turns connect
the cycles together.
\hfill\end{proof}

By strengthening the lower bound, we can achieve the following 
approximation of an optimal tour of length {\sc opt}:

\begin{theorem}\label{th:euler}
There is an $O(n\log n)$ (or $O(N)$) algorithm 
that finds a tour of turn cost at most $\frac{6}{5} \cdot${\sc opt}.
\end{theorem}

\begin{proof}
By applying Theorem~\ref{th:strong}, we get an optimal
boundary cycle cover. There are three observations that
lead to the claimed stronger results.

(1) For a thin polygon, extracting the collinear strips
can be performed in strongly polynomial time $O(n\log)$ 
(or weakly polynomial time $O(N)$).

(2) For an Eulerian thin polygon, no vertices in $G_{\overline{P}}$ remain
unmatched after repeatedly applying Lemma~\ref{le:local}.
Instead, we get an optimal cycle cover right away.
This cycle cover can be merged into one connecting tour by
merging at pixels where two cycles cross each other:
Let the optimal cycle cover be composed of $c$ disjoint cycles, 
where $c \geq 1$.  Let $t$ be the cost of the optimal cycle cover.
At each phase of the cycle-merging algorithm, two cycles are merged 
into one.  Therefore, the algorithm finds a solution having cost 
$t + 2 \cdot (c-1)$.  

(3) We can strengthen the lower bound on an optimal tour
as follows. Consider (for $c>1$) a lower bound on the cost 
of the optimal solution.
Just like in the proof of Theorem~\ref{thin hard},
all turns in a cycle cover are forced by convex corners of the polygon,
implying that any solution must contain these $t$ turns.
In addition, turning from one cycle into another incurs a
crossing cost of at least one turn; thus, we get a lower bound
of $t + c$.
Observe that there are at least $4$ turns per cycle so that 
$t \geq 4 c$.  Therefore, 
$\frac{t+2\cdot(c-1)}{t+c}\leq \frac{t+2c}{t+c}\leq \frac{6}{5}$.
\hfill\end{proof}

\subsubsection{Milling Arbitrary Thin Orthogonal Polygons}

Now we consider the case of general thin polygons. 
For any odd-degree vertex, and any feasible solution, 
some edges may have to be traversed multiple times.
As in Corollary~\ref{co:thin}, we can apply 
Theorem~\ref{th:strong} to achieve a minimum-cost
cycle cover and merge them into a tour. 
Using a more refined analysis, we can use this to
obtain a $4/3$-approximation algorithm for finding an 
minimum-cost tour.

\begin{theorem}
\label{th:4/3}
For thin orthogonal milling, we can compute a tour of turn cost at most
$\frac{4}{3}\cdot$\mbox{\sc opt} in time $O(n^3)$,
where {\sc opt} is the cost of an optimal tour. 
\end{theorem}

\begin{proof}
We start by describing how to merge the cycles into one connected tour.

\begin{enumerate}
\item find an optimal cycle cover as provided by 
Theorem~\ref{th:strong};
\item repeat until there is only one cycle in the cycle cover:
\begin{itemize}
\item If there are any two cycles that can be merged without
any extra cost (by having partially overlapping collinear edges), 
perform the merge.
%\comment{REFEREE 2:     Perhaps you could say more explicitly what this means.  I surmise that it means that they share a common edge.}
%xxx Yes, changed!
\item Otherwise,
\begin{itemize}
\item Find a vertex at which two cycles cross each other.
\item Modify the vertex to incorporate at most two additional
turns, thereby connecting the two cycles.
\end{itemize}
\end{itemize}
\end{enumerate}

Now we analyze the performance of our algorithm.
Consider the situation after extracting the cost zero
matching edges from Lemma~\ref{le:local}. This
already yields a set $K$ of cycles, obtained
by only turning at pixels of degree two
that force a turn.  
%\comment{REFEREE 2: I assume
        %that this refers to the construction from the previous page.
        %Perhaps this could be made clearer.}
%xxx Yes, I hope I did.
Let $k$ denote the number of cycles in $K$,
and $c$ be the number of turns in $K$. Let
$P$ be the set of ``dangling'' paths at degree-one or 
degree-three pixels, and 
let $p$ be the
number of turns in $P$, including the mandatory turn for each endpoint.
Let $M$ be a minimum matching between odd-degree vertices,
and let $m$ be the number of turns in $M$.
Finally, let $O$ be the matching between odd-degree
pixels that is induced by an optimal tour, and let
$o$ be the number of turns in $O$.

First note that $P$ is a matching between odd-degree nodes.

$P$, $M$ and $O$ may connect some of the cycles in $K$.  
In $P$ a path between two
odd-degree pixels connects 
the two cycles that the two nodes belong to.  
On the other hand, a path in $M$ and $O$ between two 
odd-degree nodes can encounter several cycles
along the way, and thus it may be used to merge 
several cycles at no extra cost.

Therefore, let $j$ be the number of cycles after using $P$ for free
merging,
let $i$ be the number of components with $P$ and $M$ used for free
merging, and let $h$ be the number of components with $P$ and $O$ 
used for free merging.

Note that
\begin{equation}
1 \leq i \leq j \leq k \label{one}, 
\end{equation}
and
\begin{equation}
1 \leq h \leq j \leq k \label{two}.
\end{equation}

Now consider the number of cycles encountered by a path in the 
matching. It is not hard to see that this number cannot exceed
the number of its turns. Therefore,

\begin{equation}
m \geq k - i \geq j - i \label{three}
\end{equation}
\begin{equation}
o \geq k - h \geq j - h \label{four}
\end{equation}

If a particular matching results in $x$ components, we would need at 
least another $x$ turns to get a tour.  Thus with $O$ we need at 
least $h$ more turns.

Thus, for an optimal tour of cost {\sc opt}, we have
\begin{equation}
\mbox{\sc opt} \geq c + p + o + h. \label{five}
\end{equation}

Our heuristic method adds 
the minimum matching to $C$ and $P$ and merges the
remaining components with two turns per merge, hence
the cost {\sc heur} of the resulting tour is

\begin{equation}
\mbox{\sc heur} \geq c + p + m + 2(i-1). \label{six}
\end{equation}

Thus we get the following estimate for the approximation factor $R\geq 1$:

\begin{equation}
R \leq \frac{c + p + m + 2i}{c + p + o + h} \label{seven}
\end{equation}

Because each cycle has at least four turns, we know that
\begin{equation}
c \geq 4k \geq 4j. \label{eight}
\end{equation}

Using the fact that $c+p+m+2i\geq c+p+o+h$ (because $R\geq 1$), we see
that the ratio on the right in (\ref{seven}) gets {\em larger} if we replace
$c$ in the numerator and in the denominator by the smaller nonnegative 
value $4j$; thus,

\begin{equation}
R \leq \frac{4j + p + m + 2i}{4j + p + o + h}. \label{nine}
\end{equation}

Because $P$ is also a matching, we have
\begin{equation}
p \geq m, \label{ten}
\end{equation}
which implies that
\begin{equation}
R \leq \frac{4j + 2m + 2i}{4j + m + o + h}. \label{eleven}
\end{equation}

We also know that
\begin{equation}
m \leq o. \label{twelve}
\end{equation}
Using this, together with the fact that $R$ can be assumed to be less than 2,
we can argue that $R$ is maximal for maximal values of $m$; hence,
\begin{equation}
R \leq \frac{4j + 2o + 2i}{4j + 2o + h}. \label{thirteen}
\end{equation}
Using (\ref{four}) in (\ref{thirteen}),
we see that $R$ is maximal for minimal $o$; hence,
\begin{equation}
R \leq \frac{4j + 2(j - h) + 2i}{4j + 2(j - h) + h} = \frac{6j - 2h + 2i}{6j - h}. \label{fourteen}
\end{equation}
Using $h > o$ in (\ref{fourteen}) and the facts that 
$R<2$ and $i\leq j$, 
we get that
\begin{equation}
R \leq \frac{6j -2o + 2i}{6j - o}\leq\frac{6j + 2i}{6j}\leq\frac{4}{3}. \label{fifteen}
\end{equation}
\hfill\end{proof}

The following shows that the estimate for the performance ratio is tight.

\begin{theorem}\label{ex:4over3}
There is a class of examples
for which the estimate of 4/3 for the performance ratio of the algorithm
for thin orthogonal milling is tight.
\end{theorem}

\begin{figure}[htb]
\leavevmode
\begin{center}
\epsfig{file=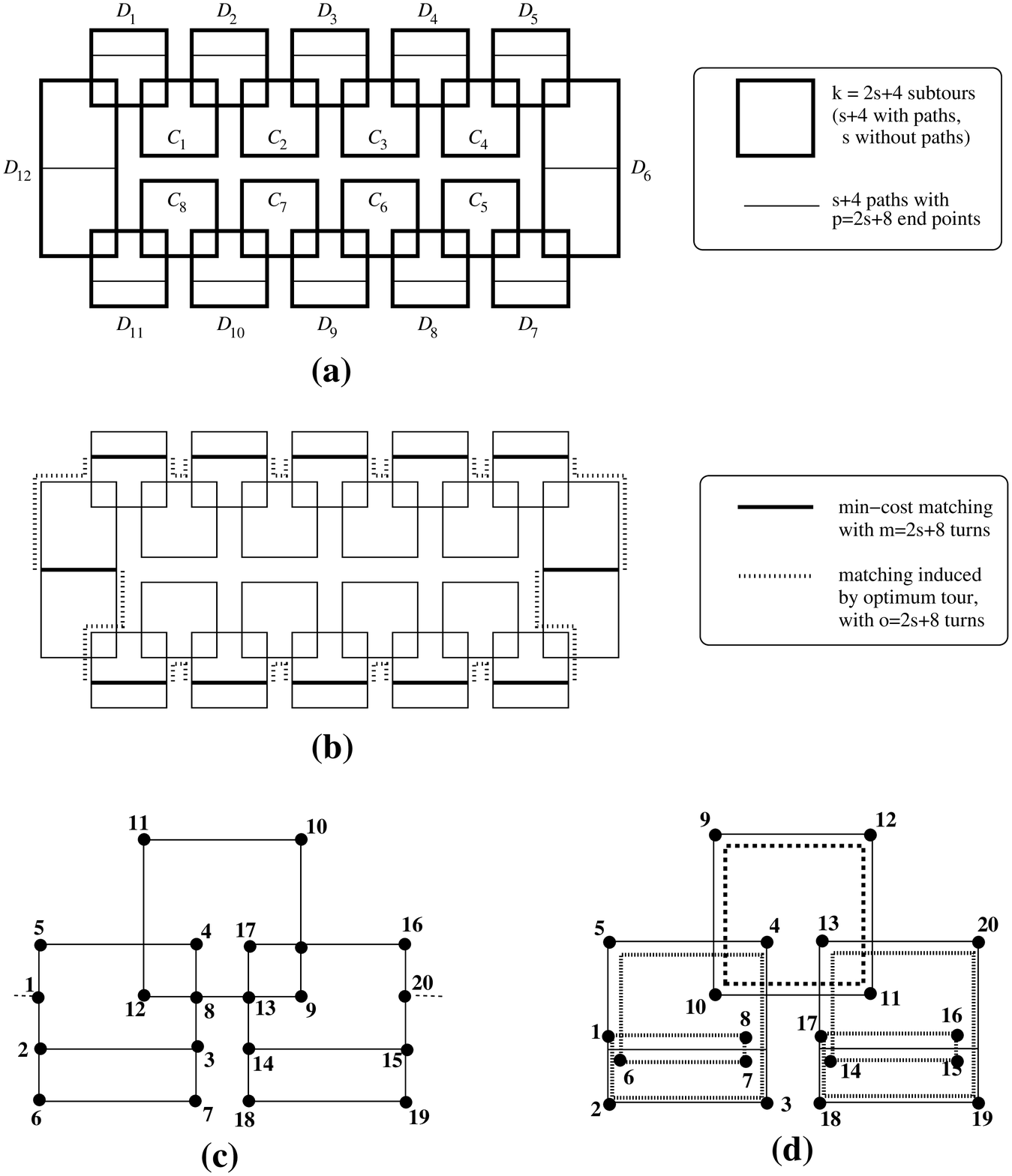,width=.90\linewidth} 
\caption{An example with performance ratio 4/3 for our heuristic. 
 (a) The structure of the example for $s=8$. 
 (b) A min-cost matching of the odd-degree vertices, 
     and the matching induced by an optimal tour. 
 (c) A portion of the optimal tour: Subtours can be merged without extra cost.
 (d) A corresponding portion of the heuristic tour: Subtours still need to be merged, which results in an additional cost of $2(k-1)=4s+6$.}
\label{fi:4over3}
\end{center}
\end{figure}

\begin{proof}
See Figure~\ref{fi:4over3}. 
%\comment{REFEREE 2: I found this hard to follow.  Could you label the
        %figure a bit better.  Also, could you indicate the optimum and
        %heuristic tours?}
%xxx DONE.
The region consists of $k=2s+4$ cycles, all with precisely 4 turns,
$s$ cycles without degree-three vertices, and $s+4$ cycles with
two degree-three vertices each. We get $c=4k=8s+16$
and $p=2(s+4)=2s+8$. Figure~\ref{fi:4over3}(b) shows
a min-cost matching of cost $m=2(s+4)=2s+8$ and one of 
cost $o=2s+8$ that is induced by an optimum tour.
As Figure~\ref{fi:4over3}(c) suggests, 
merging all cycles, odd-degree paths and matching paths is possible
without requiring any further turns, resulting in 
$\mbox{\sc opt}=c+p+o=12s+32$.
On the other hand, using the min-cost matching of cost $m$
leaves $k$ cycles that cannot be merged for free; thus, merging two cycles
at a time at a cost of 2 turns requires an additional cost of $2(k-1)=4s+6$,
for a total of $\mbox{\sc heur}=c+p+m+2(k-1)=16s+38$ turns, which gets arbitrarily
close to $\frac{4}{3} \mbox{\sc opt}$ for large $s$.
\hfill\end{proof}

Note that the argument of Theorem~\ref{th:brackets} remains valid
for this section, so the bounds on coverage and length approximation
still apply.

\subsection{PTAS}
\label{sec:ptas}

We describe a polynomial-time approximation scheme (PTAS) for the
problem of minimizing a weighted average of the two cost criteria:
length and number of turns.  Our technique is based on using the
theory of $m$-guillotine subdivisions~\cite{m-gsaps-99}, properly
extended to handle turn costs. We prove the following result:

\begin{theorem}
  Define the cost of a tour to be its length plus $C$ times the number
  of (90-degree) turns.  For any fixed $\epsilon > 0$, there is a
  $(1+\epsilon)$-approximation algorithm, with running time
  $2^{O(h)}\cdot N^{O(C)}$, for minimizing the cost of a tour for
  an integral orthogonal polygon $P$ with $h$ holes and $N$ pixels.
\end{theorem}

\begin{proof}
  Let $T^*$ be a minimum-cost tour and let $m$ be any positive
  integer.  Following the notation of \cite{m-gsaps-99}, we first
  apply the main structure theorem of that paper to claim that there
  is an $m$-guillotine subdivision, $T_G$, obtained from $T^*$ by
  adding an appropriate set of bridges ($m$-spans, which are
  horizontal or vertical segments) of total length at most ${1\over
    m}|T^*|$, with length at most $(1+{1\over m})$ times the length of
  $T^*$.  (Because $T^*$ may traverse a horizontal/vertical segment
  twice, we consider such segments to have multiplicities (1 or 2), as
  multi-edges in a graph.)
  
  We note that part of $T_G$ may lie outside the polygon $P$, because 
  the $m$-spans that we add to make $T^*$ $m$-guillotine need not lie
  within $P$.  We convert $T_G$ into a new graph by subtracting those
  portions of each bridge that lie outside of $P$.  In this way, each
  bridge of $T_G$ becomes a set of segments within $P$; we trim each
  such segment at the first and last edge of $T_G$ that is incident on
  it and call the resulting trimmed segments {\em sub-bridges}.  (Note
  that a sub-bridge may be a single point if the corresponding segment
  is incident on a single edge of $T^*$; we can ignore such trivial
  sub-bridges.)  As in the TSP method of \cite{m-gsaps-99}, we {\em
    double} the (non-trivial) sub-bridges: We replace each sub-bridge
  by a pair of coincident segments, which we ``inflate'' slightly to
  form a degenerate loop, such that the endpoints of the sub-bridge
  become vertices of degree four, and the endpoints of each edge
  incident on the interior of the sub-bridge become vertices of degree
  three (which occur in pairs).  Refer to Figure~\ref{fig:ptas-mspan}.
  We let $T_G'$ denote the resulting graph.  Now $T_G'\subset P$, and,
  because $T_G'$ is obtained from $T^*$, a tour, we know that the number
  of odd-degree vertices of $T_G'$ that lie on any one sub-bridge is
  {\em even} (the degree-three vertices along a sub-bridge come in
  pairs).

\begin{figure}[htb]
\leavevmode
\begin{center}
\epsfig{file=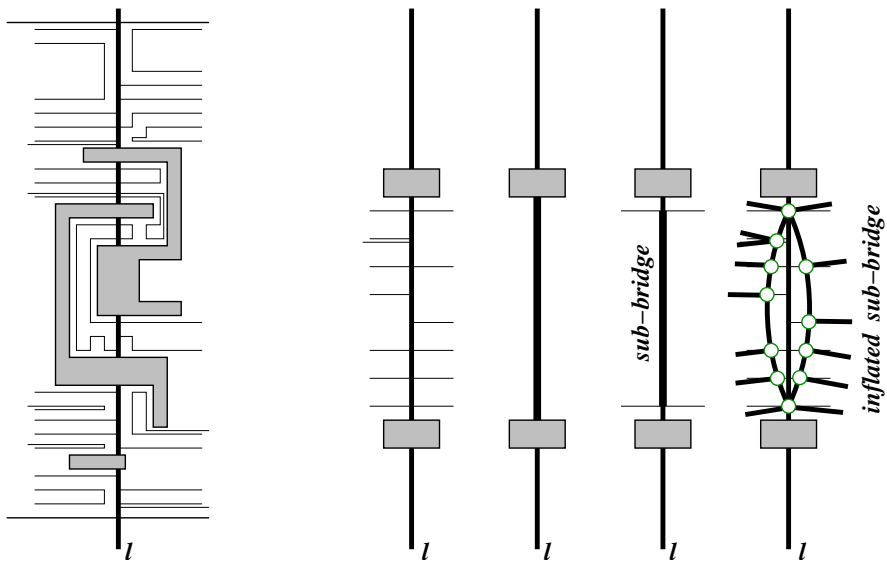,width=.90\linewidth} 
\caption{Definitions of sub-bridges and the graph $T_G'$.
A vertical line $l$ is shown, which defines a cut.  The slice of
an optimal solution along $l$ is shown, with thinner lines drawn 
along the edges of the solution (which is not intended to
be an accurate instance of an optimal solution, but is drawn to
illustrate some of the possible cases).  Also shown is an enlargement
of one portion of the cut $l$, showing a segment of the $m$-span between
two portions of the boundary of $P$, the trimmed segment that forms
the sub-bridge, and the portion of the resulting graph $T_G'$ in the
vicinity of the inflated sub-bridge.}
\label{fig:ptas-mspan}
\end{center}
\end{figure}
  
The cost of the optimal solution $T^*$ is its length, $|T^*|$, plus
$C$ times the number of its vertices.  We consider the cost of $T_G'$
to be also its (Euclidean) length plus $C$ times the number of its
vertices.  Because the number of vertices on the sub-bridges is
proportional to their total length, and each edge multiplicity is at
most two, we see that the cost of $T_G'$ is $O((1+C)/m)\cdot |T^*|$
greater than the optimal cost, i.e., the cost of $T^*$.

%REFEREE 2: A figure would help here.
% Is the idea to create something like a trapezoidal map for the
%  holes?

In order to avoid exponential dependence on $n$ in our algorithm,
we need to introduce a subdivision of $P$ that allows us to consider
the sub-bridges along an $m$-span to be grouped into a relatively
small ($O(h)$) number of classes.  
We now describe this subdivision of~$P$.

By standard plane sweep with a vertical line, we partition $P$ into
rectangles, using vertical chords, according to the vertical
decomposition; see Figure~\ref{fig:ptas-corridors}(a).  We then
decompose $P$ into a set of $O(h)$ regions, each of which is either a
``junction'' or a ``corridor.''  This decomposition is analogous to
the corridor structure of polygons that has been utilized in computing
shortest paths and minimum-link separators (see,
e.g.,~\cite{km-eaesp-88,kmm-eaesp-97,ms-sapo-95}), with the primary
difference being that we use the vertical decomposition into
rectangles, rather than a triangulation, as the basis of the
definition.  Consider the (planar) dual graph, ${\cal G}$, of the
vertical partition of $P$; the nodes of ${\cal G}$ are the rectangles,
and two nodes are joined by an edge if and only if they are adjacent.
We now define a process to transform the vertical decomposition into
our desired decomposition.  First, we take any degree-1 node of ${\cal
  G}$ and delete it, along with its incident edge; in the vertical
decomposition, we remove the corresponding vertical chord (dual to the
edge of ${\cal G}$ that was deleted).  We repeat this process, merging
a degree-1 region with its neighbor, until there are no degree-1 nodes
in ${\cal G}$.  At this stage, ${\cal G}$ has $h+1$ faces and all
nodes are of degree 2 or more.  Assume that $h\geq 2$ (the case $h\leq
1$ is easy); then, not all nodes are of degree 2, implying that there
are at least two higher-degree nodes.  Next, for each pair of adjacent
degree-2 nodes, we merge the nodes, deleting the edge between them,
and removing the corresponding vertical chord separating them in the
decomposition.  The final dual graph ${\cal G}$ has nodes of two
types: those that are dual to regions of degree 2, which we call the
{\em corridors}, and those that are dual to regions of degree greater
than 2, which we call the {\em junctions}.  Each corridor is bounded
by exactly two vertical chords, together with two portions of the
boundary of $P$.  (These two portions may, in fact, come from the same
connected component of the boundary of $P$.)  Each of the $h$ bounded
faces of ${\cal G}$ contains exactly one of the holes of $P$.  Refer
to Figure~\ref{fig:ptas-corridors}.

\begin{figure}[htb]
\leavevmode
\begin{center}
\epsfig{file=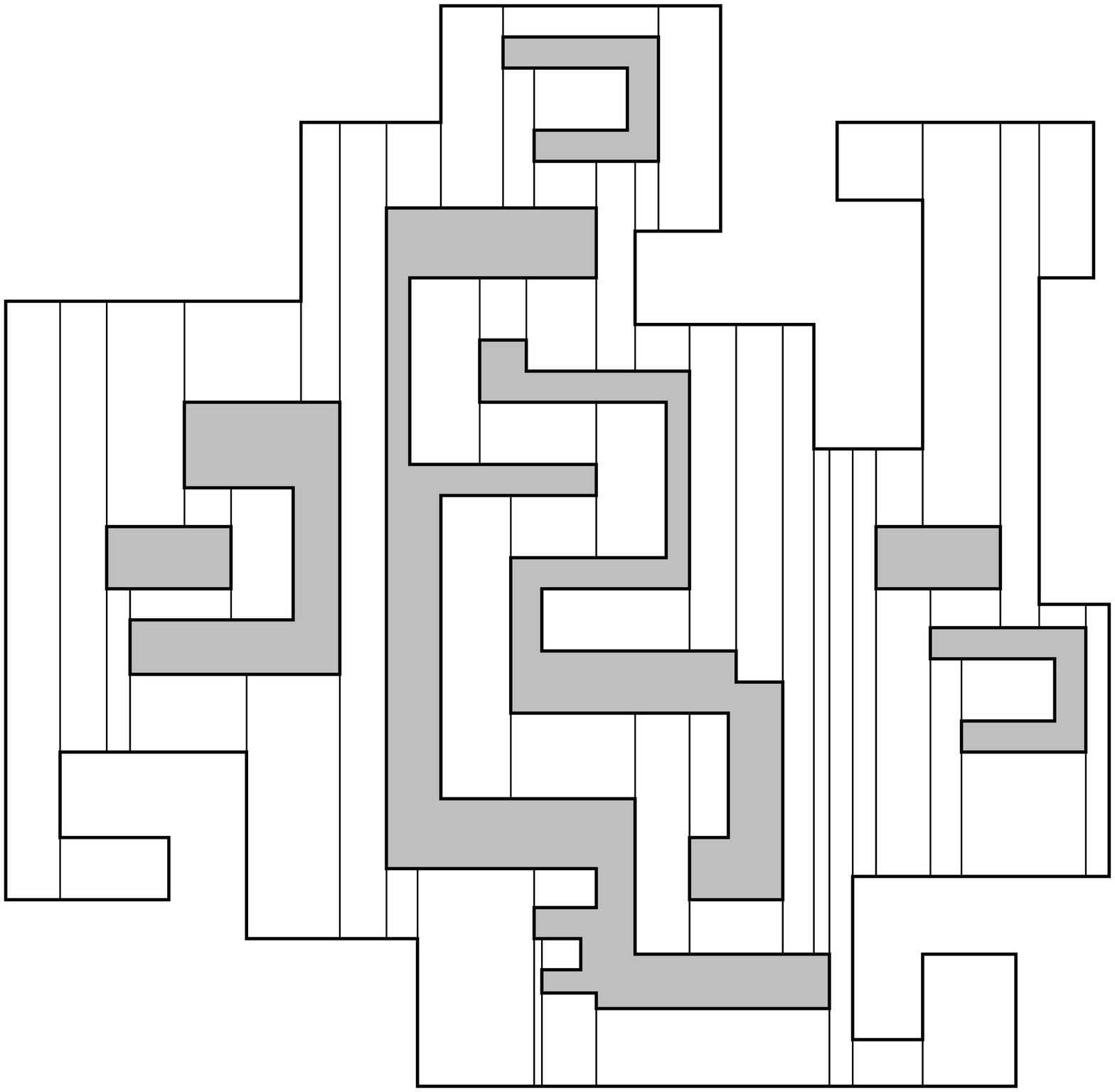,width=.45\linewidth} \hfill
\epsfig{file=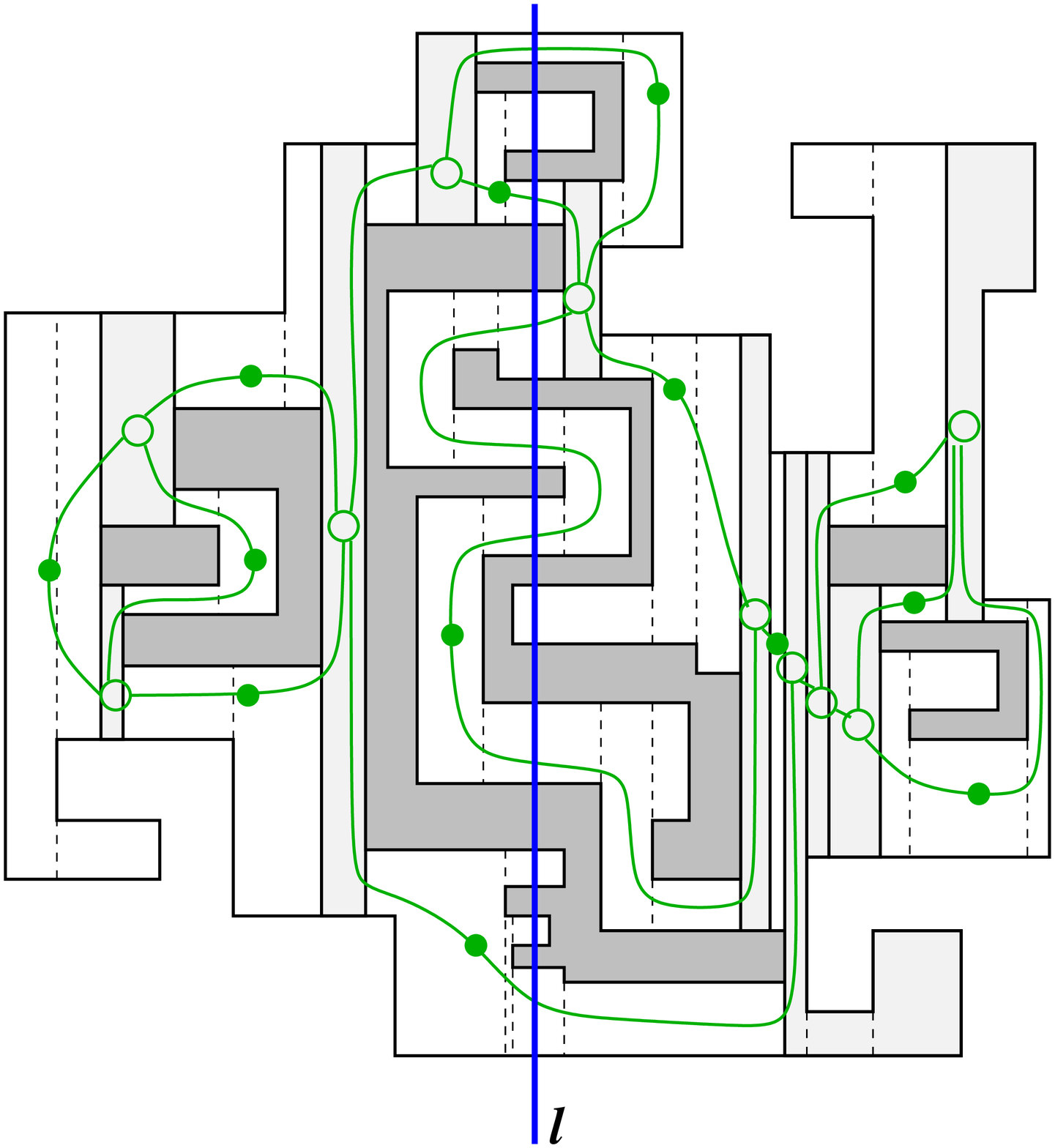,width=.45\linewidth} 
\centerline{\hfill (a) \hfill\hfill (b) \hfill}
\caption{(a). The vertical decomposition of $P$ (holes in dark grey).  (b). The decomposition of
$P$ into junctions (light grey) and corridors, with the 
dual graph overlaid.  The nodes of the dual graph are shown hollow
for junctions and shown as smaller solid disks for the
corridors.}
\label{fig:ptas-corridors}
\end{center}
\end{figure}

Let ${\cal V}$ denote the decomposition of $P$ just described; there
is an analogous horizontal partition, ${\cal H}$, of $P$ into $O(h)$
regions.  The vertical sub-bridges of a vertical bridge are
partitioned into $O(h)$ {\em classes} according to the identity of the
region, $\tau$, that contains the sub-bridge in the vertical
decomposition ${\cal V}$.  A sub-bridge intersecting a region $\tau$
of ${\cal V}$ is called {\em separating} if it separates some pair of
vertical chords on the boundary of $\tau$; it is {\em trivial}
otherwise.  (Corridor regions have only two vertical chords on their
boundary, while junctions may have several, as many as $\Theta(h)$ in
degenerate cases.)  

First, consider a corridor region $\tau$ in ${\cal
  V}$, and let $a$ and $b$ denote the two vertical chords that bound
it. An important observation regarding sub-bridge classes in corridors
is this: The parity of the number of times a tour crosses $a$ must be
the same as the parity of the number of times a tour crosses $b$.  The
consequence is that we can specify the parity of the number of
incidences on {\em all} separating sub-bridges of a given corridor
class by specifying the parity of the number of incidences on a single
sub-bridge of the class; the trivial sub-bridges always have an even
parity of crossing.  

Now consider a junction region $\tau$ in ${\cal V}$.  Because, in the
merging process that defines ${\cal V}$, we never merge a degree-2
region with a higher-degree region, we know that $\tau$ consists of a
single high-degree ($>2$) rectangle, $R_\tau$, from the original
vertical decomposition, together with possibly many other rectangles
that form a ``pocket'' attached to $R_\tau$, corresponding to a tree
in the dual graph (so that removal of degree-1 nodes leads to a
collapse of the tree and a merging of the pocket to $R_\tau$).  The
consequence of this observation is that there can be at most one
separating sub-bridge in a junction class. (There may be several
trivial sub-bridges.)

  Our algorithm applies dynamic programming to obtain a minimum-cost
  $m$-guillo\-tine subdivision, $T_G^*$, from among all those
  $m$-guillotine subdivisions that have the following additional
  properties:
\begin{description}
\item[(1)] It consists of a union of horizontal/vertical segments,
  having half-integral coordinates, within~$P$.
\item[(2)] It is connected.
\item[(3)] It {\em covers} $P$, in that the center of every pixel of $P$ is
  intersected by an edge of the subdivision.
\item[(4)] It is a {\em bridge-doubled} $m$-guillotine subdivision, so
  that every (non-trivial) sub-bridge of an $m$-span appears twice (as
  a multi-edge).
\item[(5)] It interconnects the sub-bridges in each of a specified
  partition of the classes of sub-bridges.
\item[(6)] It obeys a parity constraint on each of the $O(h)$ classes
  of sub-bridges: The number of edges of the subdivision incident on
  each separating sub-bridge of the class corresponding to a region
  $\tau$ is even or odd, according to the specified parity for~$\tau$.
\end{description}

A subproblem in the dynamic programming algorithm is specified by a
rectangle, having half-integral coordinates, together with {\em
  boundary information} associated with the rectangle, which specifies
how the subdivision within the rectangle interacts with the
subdivision outside the rectangle.  The boundary information includes
(a) $O(m)$ attachment points, where edges meet the boundary at points
other than along the $m$-span; (b) the multiplicity (1 or 2) of each
attachment point, and the interconnection pattern (if any) of adjacent
attachment points along the rectangle boundary; (c) the endpoints of a
bridge on each side of the rectangle (from which one can deduce the
sub-bridges); (d) interconnection requirements among the attachment
points and the classes of sub-bridges; and (e) a parity bit for each
class of sub-bridge, indicating whether an even or an odd number of
edges should be incident to the separating sub-bridges of that class.
There are $N^{O(m)}\cdot 2^{O(h)}$ subproblems.  At the base of the
dynamic programming recursion are rectangles of constant size (e.g.,
unit squares).

The optimization considers each possible cut (horizontal or vertical,
at half-integral coordinates) for a given subproblem, together with
all possible choices for the new boundary information along the cut,
and minimizes the total resulting cost, adding the costs of the two
resulting subproblems to the cost of the choices made at the cut
(which includes the length of edges added, plus $C$ times the number
of vertices added).

Once an optimal subdivision, $T_G^*$, is computed, we can recover a
valid tour from it by extracting an Eulerian subgraph obtained by
removing a subset of the edges on each doubled sub-bridge.  The parity
conditions imply that such an Eulerian subgraph exists.  An Eulerian
tour on this subgraph is a covering tour, and its cost is at most
$O(C/m)\cdot |T^*|$ greater than optimal.  For any fixed $\epsilon>0$,
we set $m=\lceil C/\epsilon\rceil$, resulting in a
$(1+\epsilon)$-approximation algorithm with running time
$O(N^{O(C/\epsilon)}\cdot 2^{O(h)}$.
The techniques of \cite{m-gsaps-99}, which use ``grid-rounded''
$m$-guillotine subdivisions, can be applied to reduce the exponent on
$N$ to a term, $O(C)$, independent of~$m$.  \hfill\end{proof}

\paragraph{Remarks}
It should be possible to apply a variant of our methods to obtain a
PTAS that is polynomial in $n$ (versus $N$), with a careful
consideration of implicit encodings of tours.  We note that our result
relies on effectively ``charging off'' turn cost to path length, because 
the objective function is a linear combination of the two costs (turns
and length).  We have not yet been able to give a PTAS for minimizing
only the number of turns in a covering tour; this remains an
intriguing open problem.

\old{
\subsection{PTAS}
\label{sec:ptas}

We describe a polynomial-time approximation scheme (PTAS) for the
problem of minimizing a weighted average of the two cost criteria:
length and number of turns.  Our technique is based on using the
theory of $m$-guillotine subdivisions~\cite{m-gsaps-99}, properly
extended to handle turn costs. We prove the following result:

\begin{theorem}
  Define the cost of a tour to be its length plus $C$ times the number
  of (90-degree) turns.  For any fixed $\epsilon > 0$, there is a
  $(1+\epsilon)$-approximation algorithm, with running time
  $2^{O(h)}\cdot N^{O(C)}$, for minimizing the cost of a tour for
  an integral orthogonal polygon $P$ with $h$ holes and $N$ pixels.
\end{theorem}

\begin{proof}
  Let $T^*$ be a minimum-cost tour and let $m$ be any positive
  integer.  Following the notation of \cite{m-gsaps-99}, we first
  apply the main structure theorem of that paper to claim that there
  is an $m$-guillotine subdivision, $T_G$, obtained from $T^*$ by
  adding an appropriate set of bridges ($m$-spans, which are
  horizontal or vertical segments) of total length at most ${1\over
    m}|T^*|$, with length at most $(1+{1\over m})$ times the length of
  $T^*$.  (Because $T^*$ may traverse a horizontal/vertical segment
  twice, we consider such segments to have multiplicities, as
  multi-edges in a graph.)
  
  We note that part of $T_G$ may lie outside the polygon $P$, because 
  the $m$-spans that we add to make $T^*$ $m$-guillotine need not lie
  within $P$.  We then convert $T_G$ into a new graph by removing
  those portions of each bridge that lie outside of $P$.  In this way,
  each bridge of $T_G$ becomes a set of segments within $P$; we trim
  each such segment at the first and last edge of $T_G$ that crosses
  it and call the resulting trimmed segments {\em sub-bridges}.  (Note
  that a sub-bridge may be a single point if the corresponding segment
  is incident on a single edge of $T^*$; we can ignore such trivial
  sub-bridges.)  As in the TSP method of \cite{m-gsaps-99}, we {\em
    double} the (non-trivial) sub-bridges: We replace each sub-bridge
  by a pair of coincident segments, which we ``inflate'' slightly to
  form a degenerate loop, such that the endpoints of the sub-bridge
  become vertices of degree four, and the endpoints of each edge incident
  on the interior of the sub-bridge become vertices of degree three (which
  occur in pairs).  We let $T_G'$ denote the resulting graph.  Now
  $T_G'\subset P$, and, because $T_G'$ is obtained from $T^*$, a tour,
  we know that the number of odd-degree vertices of $T_G'$ that lie on
  any one sub-bridge is {\em even} (the degree-three vertices along a
  sub-bridge come in pairs).
  
  The cost of the optimal solution $T^*$ is its length, $|T^*|$, plus
  $C$ times the number of its vertices.  We consider the cost of
  $T_G'$ to be also its (Euclidean) length plus $C$ times the number
  of its vertices.  Because each sub-bridge is of integral length, and
  each edge multiplicity is at most two, we see that the cost of
  $T_G'$ is at most $O(C/m)\cdot |T^*|$ greater than the optimal cost
  (the cost of $T^*$).

  We need one more piece of notation.  
\comment{REFEREE 2: A figure would help here.
        Is the idea to create something like a trapezoidal map for the
        holes?}
Because polygon $P$ has $h$
  holes, it can be partitioned using vertical chords into $O(h)$
  subregions, each of which is bounded by two chords and two connected
  portions of the boundary of $P$ (one portion from each of two holes,
  or from the outer boundary of $P$).  Let ${\cal V}$ denote this
  decomposition of $P$.  The vertical sub-bridges of a vertical bridge
  are partitioned into $O(h)$ {\em classes} according to the identity
  of the region, $\tau$, that contains the sub-bridge in the vertical
  decomposition ${\cal V}$.  Similarly, a horizontal partition, ${\cal
    H}$, of $P$ into $O(h)$ regions results in a partitioning of the
  horizontal sub-bridges of any horizontal bridge into $O(h)$ classes.
  Consider a region $\tau$ in ${\cal V}$, and let $a$ and $b$ denote
  the two vertical chords that bound it. The important observation
  regarding sub-bridge classes is this: The parity of the number of
  times a tour crosses $a$ must be the same as the parity of the
  number of times a tour crosses $b$.  The consequence is that we can
  specify the parity of the number of incidences on {\em all}
  sub-bridges of a given class by specifying the parity of the number
  of incidences on a single sub-bridge of the class.

  Our algorithm applies dynamic programming to obtain a minimum-cost
  $m$-guillo\-tine subdivision, $T_G^*$, from among all those
  $m$-guillotine subdivisions that have the following additional
  properties:
\begin{description}
\item[(1)] It consists of a union of horizontal/vertical segments,
  having integral coordinates, within~$P$.
\item[(2)] It is connected.
\item[(3)] It {\em covers} $P$, in that every pixel of $P$ is
  intersected by an edge of the subdivision.
\item[(4)] It is a {\em bridge-doubled} $m$-guillotine subdivision, so
  that every (non-trivial) sub-bridge of an $m$-span appears twice (as
  a multi-edge).
\item[(5)] It interconnects the sub-bridges in each of a specified
partition of classes of sub-bridges.
\item[(6)] It has a number of edges incident on each class of
  sub-bridge that satisfies a specified parity (even or odd)
  condition.
\end{description}

A subproblem in the dynamic programming algorithm is specified by a
rectangle, together with {\em boundary information} associated with
the rectangle, which specifies how the subdivision within the
rectangle interacts with the subdivision outside the rectangle.  The
boundary information includes (a) $O(m)$ attachment points (with their
multiplicity, 1 or 2), (b) the endpoints of a bridge on each side of
the rectangle (from which one can deduce the sub-bridges), (c)
interconnection requirements among the attachment points and the
classes of sub-bridges, and (d) a parity bit for each class of
sub-bridge, indicating whether an even or an odd number of edges
should be incident to the sub-bridges of that class.  
There are $O(N^{O(m)}\cdot 2^{O(h)})$ subproblems.

Once an optimal subdivision, $T_G^*$, is computed, we can recover a
valid tour from it by extracting an Eulerian subgraph obtained by
removing a subset of the edges on each doubled sub-bridge.  The parity
conditions imply that such an Eulerian subgraph exists.  An Eulerian
tour on this subgraph is a covering tour, and its cost is at most
$O(C/m)\cdot |T^*|$ greater than optimal.  For any fixed $\epsilon>0$,
we set $m=\lceil C/\epsilon\rceil$, resulting in a
$(1+\epsilon)$-approximation algorithm with running time
$O(N^{O(C/\epsilon)}\cdot 2^{O(h)}$.
The techniques of \cite{m-gsaps-99}, which use ``grid-rounded''
$m$-guillotine subdivisions, can be applied to reduce the exponent on
$N$ to a term, $O(C)$, independent of~$m$.  \hfill\end{proof}

\paragraph{Remarks}
It should be possible to apply a variant of our methods to obtain a
PTAS that is polynomial in $n$ (versus $N$), with a careful
consideration of implicit encodings of tours.  We note that our result
relies on effectively ``charging off'' turn cost to path length, because 
the objective function is a linear combination of the two costs (turns
and length).  We have not yet been able to give a PTAS for minimizing
only the number of turns in a covering tour; this remains an
intriguing open problem.
}

\section{Conclusion}
We have presented a variety of results for finding an optimal tour with
turn cost. Many open problems remain. The most curious seems to be the
following, which highlights the difference between turn cost and length,
as well as the difference between a cycle cover and a 2-factor:

\begin{problem}
\label{pr:cc}
  What is the complexity of finding a minimum-turn cycle cover in 
  a grid graph?
\end{problem}

This problem has been open for several years now; in fact, it is 
Problem \#\,53 on the well-known list \cite{topp},
known as ``The Open Problems Project''.
    Finding a minimum weighted turn cycle cover is known
    to be NP-hard for a set of points in the plane~\cite{ackms-amtsp-97};
    however, the proof uses the fact that there are more than two directions
    for the edges in a convex cycle. While we tend to believe that
    Problem~\ref{pr:cc} may have the answer ``NP-complete'',
    a polynomial solution 
   would immediately lead to a 1.5-approximation for the orthogonal case,
    and a $(1+{2\over 3})$-approximation for the general case.

For various optimization problems dealing with geometric regions, 
there is a notable difference in complexity between a
region with holes, and a {\em simple} region that does not have any holes.
(In particular, it can be decided in polynomial time whether a given grid
graph without holes has a Hamiltonian cycle~\cite{ul-hcsgg-97}, 
even though the complexity of the TSP on these graphs is still open.)
Our NP-hardness proof makes strong use of holes; furthermore, the
complexity of the PTAS described above is exponential in the number
of holes. This raises the following natural question.

\begin{problem}
\label{pr:simple}
  Is there a polynomial-time algorithm for exactly computing a
  minimum-turn covering tour for {\em simple} orthogonal polygons?
\end{problem}

It may be possible to improve the performance of some of our 
approximation algorithms. In particular, the following
remains unclear.

\begin{problem}
\label{pr:375}
  Is the analysis of the 3.75-approximation algorithm tight?
\end{problem}

We believe that it may be possible to improve the factor.
We also believe that there is room for improvement in approximating
nonintegral orthogonal milling, in particular by improving the
cost of finding a strip cover.

\begin{problem}
\label{pr:strip.fr}
  What is the complexity of computing a minimum strip cover in nonintegral orthogonal polygons?
\end{problem}

An important tool for our approximation algorithms is a strip cover
of small cost; finding a strip cover remains a possible approach
even if strips may be parallel to more than two directions. 
This is closely related to other decomposition problems;
see \cite{k-pd-00} for a survey.

\begin{problem}
\label{pr:strip.nonorth}
  What is the complexity of computing minimum strip covers in nonorthogonal polygons?
\end{problem}

The answer may very well be ``NP-hard, even for three directions'':
Hassin and Megiddo~\cite{hm-aahos-91} show that the problem of hitting a set of points
with a minimum number of lines with three slopes is hard. However, their
proof constructs a disconnected set of grid points and cannot be applied
directly to milling problems.
In any case, even an approximation would be of interest; in particular,
if it achieves the following property:
 
\begin{problem}
\label{pr:strip.direct}
  Is there a strip cover approximation algorithm for $d$ directions whose performance ratio is independent of $d$?
\end{problem}

This would imply a positive result for a special case of the following,
even more general problem.

\begin{problem}
\label{pr:approx.arbit}
  Can one obtain approximation algorithms for unrestricted directions in an arbitrary polygonal domain, and an appropriately shaped cutter?
\end{problem}

\section*{Acknowledgments}
We are obliged to Valentin Polishchuk for a very thorough list
of suggestions, and thank three anonymous referees for various comments
that helped to improve the presentation of the paper.
We thank Regina Estkowski for helpful discussions.
This research was partially supported by the National Science
Foundation (CCR-9732221, CCR-0098172) and by
grants from Bridgeport Machines, HRL Laboratories, ISX Corporation, Metron Aviation, NASA (NAG2-1325), Northrop-Grumman, Sandia National Labs, 
Seagull Technology, and Sun Microsystems. Various parts of this work were
done while S\'andor Fekete visited Stony Brook, with partial
support by DFG travel grants Fe407. Saurabh Sethia participated in this
research while affiliated to the Department of Applied Mathemathematics 
and Statistics, SUNY Stony Brook.

\bibliography{refs,../submit/geom}
\bibliographystyle{siam}

\end{document}